\documentstyle[multicol,aps,rmp,amsfonts,graphics,epsfig]{revtex}
\begin{document}
\title{Saturation of electrical resistivity}                              
\author{O. Gunnarsson$^{(1)}$, M. Calandra$^{(2)}$ and J.E. Han$^{(3)}$}
\address{${}^{(1)}$ Max-Planck-Institut f\"ur Festk\"orperforschung, 
Postfach 800665, D-70506 Stuttgart, Germany \\
${}^{(2)}$Laboratoire de Min\'eralogie-Cristallographie, case 115, 
4 Place Jussieu, 75252, Paris cedex 05, France \\
${}^{(3)}$Department of Physics, The Pennsylvania State University,
University Park, PA 16802-6300}

\maketitle

\begin{abstract}
Resistivity saturation is observed in many metallic systems with  
large resistivities, i.e., when the resistivity has reached a critical
value, its further increase with temperature is substantially reduced. 
This typically happens when the apparent mean free path is comparable 
to the interatomic separations - the Ioffe-Regel condition. Recently, 
several exceptions to this rule have been found. Here, we review 
experimental results and early theories of resistivity saturation.
We then describe more recent theoretical work, addressing cases both 
where the Ioffe-Regel condition is satisfied and where it is violated. 
In particular we show how the (semiclassical) Ioffe-Regel condition 
can be derived quantum-mechanically under certain assumptions about 
the system and why these assumptions are violated for high-$T_c$
cuprates and alkali-doped fullerides.
\end{abstract}

\begin{multicols}{2}
\tableofcontents
\section{Introduction}\label{sec:a}
The electrical resistivity, $\rho$, of metals is usually calculated in
the Boltzmann theory, where the electrons are treated semiclassically.
An electron is assumed to move with a wave vector 
${\bf k}$ between scattering events, caused by phonons,  
other electrons, impurities or some other disorder. The average 
distance an electron moves between two scattering events is    
the mean free path $l$. Assuming a three-dimensional system and a 
spherical Fermi surface with one sheet, $\rho$ can be expressed 
in terms of $l$ as (see Appendix A) 
\begin{equation}\label{eq:1}
\rho={3\pi^2 \hbar\over e^2k_F^2 l},
\end{equation}
where $k_F$ is the Fermi wave vector. Alternatively, if the resistivity 
is known experimentally, an apparent mean free path can be determined 
from Eq.  (\ref{eq:1}). In the following, Eq. (\ref{eq:1}) will
be used in the latter way to determine $l$, which then only depends 
on the experimental resistivity and the density of conduction electrons 
(giving $k_F$).

\begin{figure}[t]
\centerline{
\rotatebox{-90}{\resizebox{2.0in}{!}{\includegraphics{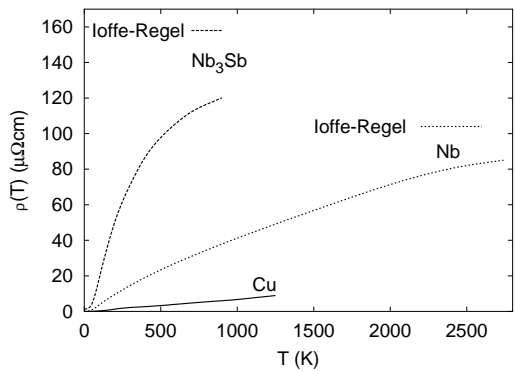}} }}
\caption[]{\label{fig:1}Resistivity of Cu, Nb$_3$Sb (Fisk and Webb, 
1976) and Nb (Abraham and Deviot, 1972). The figure also shows the the 
Ioffe-Regel (Ioffe and Regel, 1960) saturation resistivities of Nb$_3$Sb
and Nb, setting the mean free path $l$ in Eq. (\ref{eq:1}) equal to the  
distance between the Nb atoms. The corresponding value for Cu, 260
$\mu\Omega$cm, falls outside the figure. The figure illustrates that 
for Nb$_3$Sb and Nb the resistivity saturates roughly as predicted by 
the Ioffe-Regel criterion, while $\rho(T) \sim T$ for Cu at large $T$.}
\end{figure}

For a good metal like Cu, $l$ is of the order of hundreds or even 
thousands of \AA. In the semiclassical Boltzmann theory and for $T$ 
larger than a fraction of a typical phonon energy, the resistivity 
due to phonon scattering grows linearly with $T$. This is shown           
for Cu in Fig. \ref{fig:1}. The apparent mean free path is then reduced 
correspondingly. For a good metal, however, $l$ is much larger than the 
separation $d$ of the atoms, even at the melting temperature. 

For certain metals, in particular many transition metals and 
transition metal compounds, the resistivity behaves in a completely 
different way. This was emphasized by Fisk and Webb (1976),     
who studied the A15 compounds Nb$_3$Sb (see Fig. \ref{fig:1}).
Similar results had earlier been obtained by Woodard and Cody (1964)
for Nb$_3$Sn and by several other groups for other compounds. 
For small $T$§, $\rho(T)$ grows much faster than for Cu and $l$     
becomes comparable to $d$ already for $T$ of the order of several 
hundred K. At this point, the slope of the resistivity curve is
substantially reduced, and $l$ stays comparable to 
$d$ for experimentally accessible values of $T$. This is referred to 
as resistivity saturation. It describes the situation where $\rho(T)$ 
grows much slower than $\rho(T)\sim T$, predicted by the Boltzmann 
equation, but it does not necessarily mean that $\rho(T)$ becomes 
a constant. 

Ioffe and Regel (1960) pointed out that the semiclassical
theory makes no sense if $l<d$, and we refer to $l \gtrsim d$
as the Ioffe-Regel condition. Inserting $l=d$ in Eq. (\ref{eq:1}) 
gives the Ioffe-Regel resistivity, shown in Fig. \ref{fig:1}. 
In a semiclassical picture, it is natural that $l$ cannot be much 
smaller than $d$, since one may expect that an electron at most is
scattered at every atom. Saturation is then expected 
when $l \sim d$. A semiclassical theory, however,  breaks down when 
$l\sim d$, since the uncertainty in the ${\bf k}$-vector
of an electron is comparable to the size of the Brillouin
zone. A semiclassical theory cannot therefore explain 
saturation. Nevertheless, as discussed in Sec. \ref{sec:b}, in the 
1970's and early 1980's a large number of metals were found which showed 
saturation when $l\sim d$, and the behavior seemed to be universal. 
During this time, much theoretical work was performed to explain 
saturation. No consensus was reached, however, and the interest 
turned to other problems.                  

\begin{figure}[t]
\centerline{
\rotatebox{-90}{\resizebox{2.3in}{!}{\includegraphics{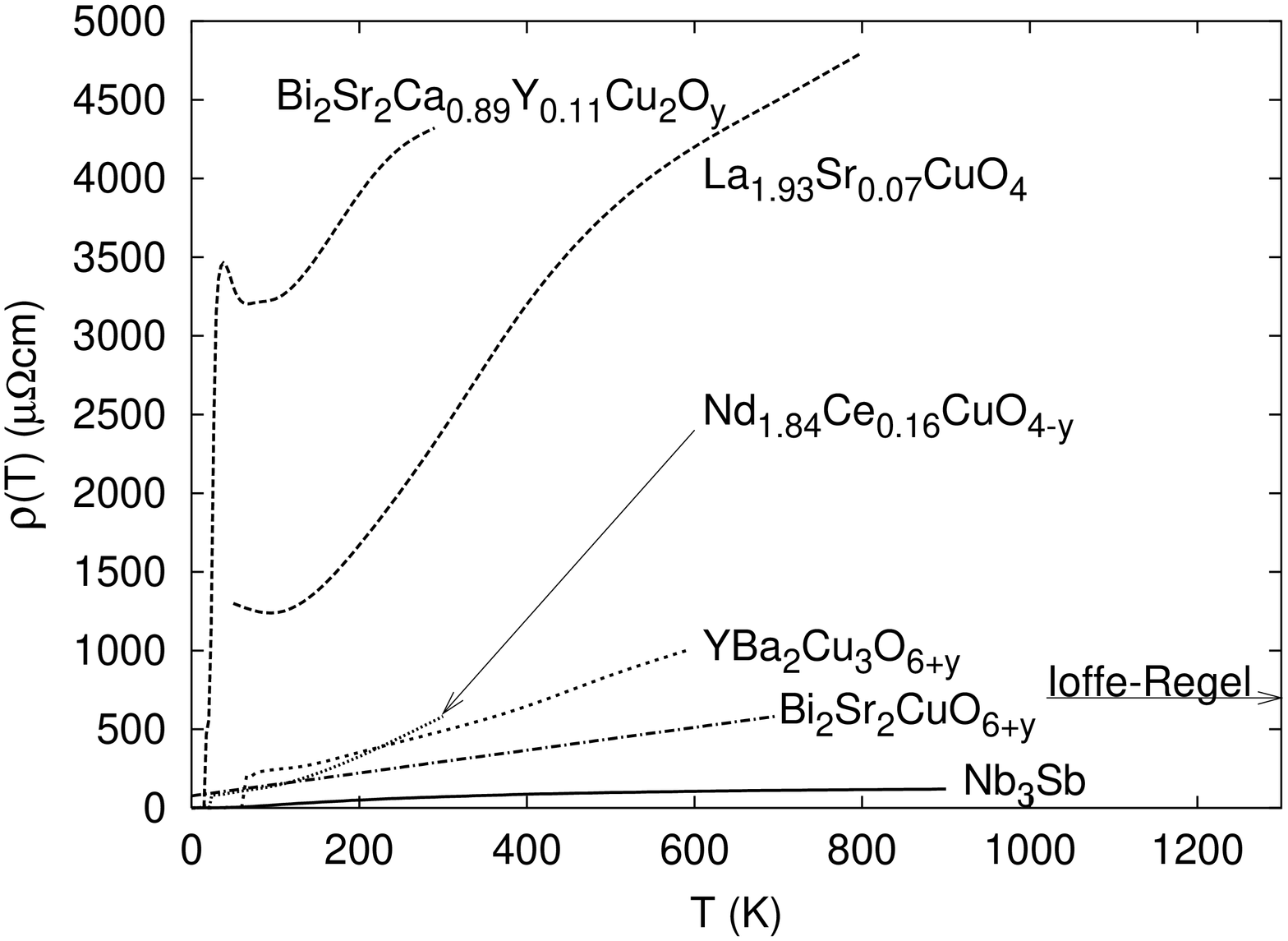}} }}
\caption[]{\label{fig:2}Resistivity of
Bi$_2$Sr$_2$Ca$_{1-x}$Y$_x$Cu$_2$O$_{8+y}$  ($T_c=30$ K) (Wang 
{\it et al.}, 1996ab), La$_{1.93}$Sr$_{0.07}$CuO$_4$  (Takagi 
{\it et al.}, 1992), Nd$_{1.84}$Ce$_{0.16}$Cu$_{4-y}$ ($T_c=22.5$ K)
(Hikada and Suzuki, 1989), YBa$_2$Cu$_3$O$_{6+x}$ ($T_c=60$ K)
(Orenstein {\it et al.}, 1990), Bi$_2$Sr$_2$Cu$_{6+y}$ ($T_c=6.5$ K)
(Martin {\it et al.}, 1990) and Nb$_3$Sb (Fisk and Webb, 1976). 
The arrow shows the Ioffe-Regel resistivity of 
La$_{1.93}$Sr$_{0.07}$CuO$_4$. The figure illustrates 
that there is no sign of saturation at the Ioffe-Regel resistivity, 
but in some cases perhaps at much larger resistivities. Observe the 
magnitude compared with Nb$_3$Sb.}
\end{figure}

\begin{figure}[t]
\centerline{
\rotatebox{-90}{\resizebox{2.2in}{!}{\includegraphics{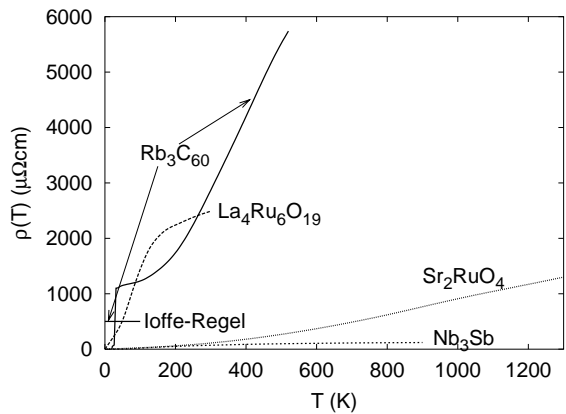}} }}
\caption[]{\label{fig:3}Resistivity of Rb$_3$C$_{60}$
(Hebard {\it et al.}, 1993) La$_4$Ru$_6$O$_{19}$ (Khalifah {\it et al.},
2001), Sr$_2$RuO$_4$ (Tyler {\it et al.}, 1998), Nb$_3$Sb (Fisk and 
Webb, 1976) and the Ioffe-Regel resistivity for Rb$_3$C$_{60}$. There 
is no sign of saturation at the Ioffe-Regel resistivity, but 
La$_4$Ru$_6$O$_{19}$ may saturate at a much larger resistivity. }
\end{figure}

The situation changed drastically in the late 1980's, when it was found 
that the high-$T_c$ cuprates behave very differently (Gurvitch and Fiory, 
1987). Fig. \ref{fig:2} shows that the resistivity for these compounds 
is typically much larger than for metals as Nb$_3$Sb, 
satisfying the Ioffe-Regel condition. As described in Appendix A, the 
Ioffe-Regel resistivity for La$_{2-x}$Sr$_x$CuO$_4$ is very large, 
about 0.7 m$\Omega$cm, due to the small carrier concentration. 
Nevertheless, the resistivities of the high-$T_c$ cuprates  
substantially exceed the Ioffe-Regel resistivity
or are comparable to this resistivity without signs
of saturation. The Ioffe-Regel condition is therefore strongly
violated. Both La$_{2-x}$Sr$_x$CuO$_4$ for small $x$ and 
Bi$_2$Sr$_2$Ca$_{1-x}$Y$_x$Cu$_2$O$_{8+y}$ show, however, signs 
of saturation, although at much larger values than the Ioffe-Regel 
resistivity. 
Fig. \ref{fig:3} shows other examples of the violation of the 
Ioffe-Regel condition, namely Rb$_3$C$_{60}$ (Hebard {\it et al.}, 
1993), La$_4$Ru$_6$O$_{19}$ (Khalifah {\it et al.}, 2001) and
Sr$_2$RuO$_4$ (Tyler {\it et al.}, 1998).

Although there are substantial uncertainties in the absolute values 
of the experimental resistivity, it is, nevertheless, clear that the 
Ioffe-Regel condition can be violated. This shows that the semiclassical 
argument for saturation is not just questionable, but actually gives 
wrong predictions for these systems. This emphasizes the need for a 
theory of saturation going beyond the semiclassical treatment of the 
electrons. Such a theory should also explain the violation of the 
Ioffe-Regel condition for the systems mentioned above. The drastic 
change in the experimental situation over the last 15 years has led 
to a renewed interest in resistivity saturation.                     

In Sec. \ref{sec:b}, we review the experimental results, and in Sec.
\ref{sec:c} early theoretical work. We then briefly describe methods
for calculating the resistivity in Sec. \ref{sec:ca}. 
Much of the theoretical work has started from 
a Boltzmann-like approach, assuming that the scattering mechanism 
is a relatively weak perturbation. In Sec. \ref{sec:d}, we describe 
an approach starting from the opposite limit of very strong scattering,
assuming that the Drude peak has been completely removed. This approach 
is based on the f-sum rule. It leads to an approximate upper limit 
$\rho_{\rm sat}$ to the resistivity, which usually has a weak $T$ 
dependence. Resistivity saturation then happens if the initial slope 
of $\rho(T)$ is so large that $\rho_{\rm sat}$ is reached at small 
values of $T$ and if $\rho_{\rm sat}$ has a weak $T$ dependence.
This approach is  applied to three models with different saturation 
behavior. In Sec. \ref{sec:e}, we describe the treatment of a model 
of weakly correlated electrons in a broad band, appropriate for many 
transition metal compounds. This model shows saturation in agreement 
with the Ioffe-Regel condition. We show how this condition can be derived 
quantum-mechanically by assuming i) noninteracting electrons and ii) $T\ll 
W$, where $W$ is the band width. In Sec. \ref{sec:f} we treat high-$T_c$ 
cuprates, for which assumption i) about noninteracting electrons is
violated. Using the $t-J$ model, we find that saturation can occur, but at  
much larger resistivities than predicted by the Ioffe-Regel condition. 
Other metals violating the Ioffe-Regel condition are discussed in 
Sec. \ref{sec:g}. In Sec. \ref{sec:h}, we describe a model of 
alkali-doped C$_{60}$, which shows no saturation. This result depends on
the intramolecular character of the phonons and on condition ii) ($T\ll W$)
being violated.
The relation to Anderson localization and Mott's minimum
conductivity is discussed in Sec. \ref{sec:i}.

\begin{figure}
\centerline{
\rotatebox{0}{\resizebox{1.7in}{!}{\includegraphics{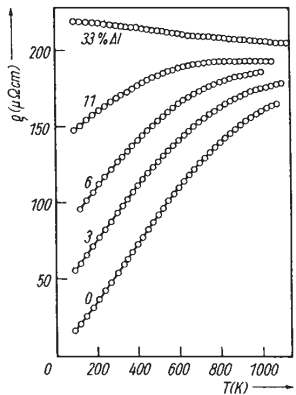}} }}
\caption[]{\label{fig:4}Resistivity of Ti$_{1-x}$Al$_x$ alloys.
The figure suggests that the saturation resistivity is 
independent of the disorder (after Mooij, 1973).}
\end{figure}
 
\section{Experimental results}\label{sec:b}
Most cases of
resistivity saturation have been found for transition metal compounds.
Mooij (1973) made one of the first observations for Ti$_{1-x}$Al$_x$ 
(see Fig. \ref{fig:4}) and several other alloys. In addition, he found 
that the positive $T$ dependence of the resistivity becomes weaker or 
even negative for strongly disordered systems. Similar results were found 
by Arko {\it et al.} (1973) and by Tsuei (1986) for other compounds. 
These results suggest that compositional disorder and disorder due to 
thermally excited phonons can have a similar effect. Other early 
examples were found by Fisk and Lawson (1973).

\begin{figure}[t]
\centerline{
\rotatebox{-90}{\resizebox{1.9in}{!}{\includegraphics{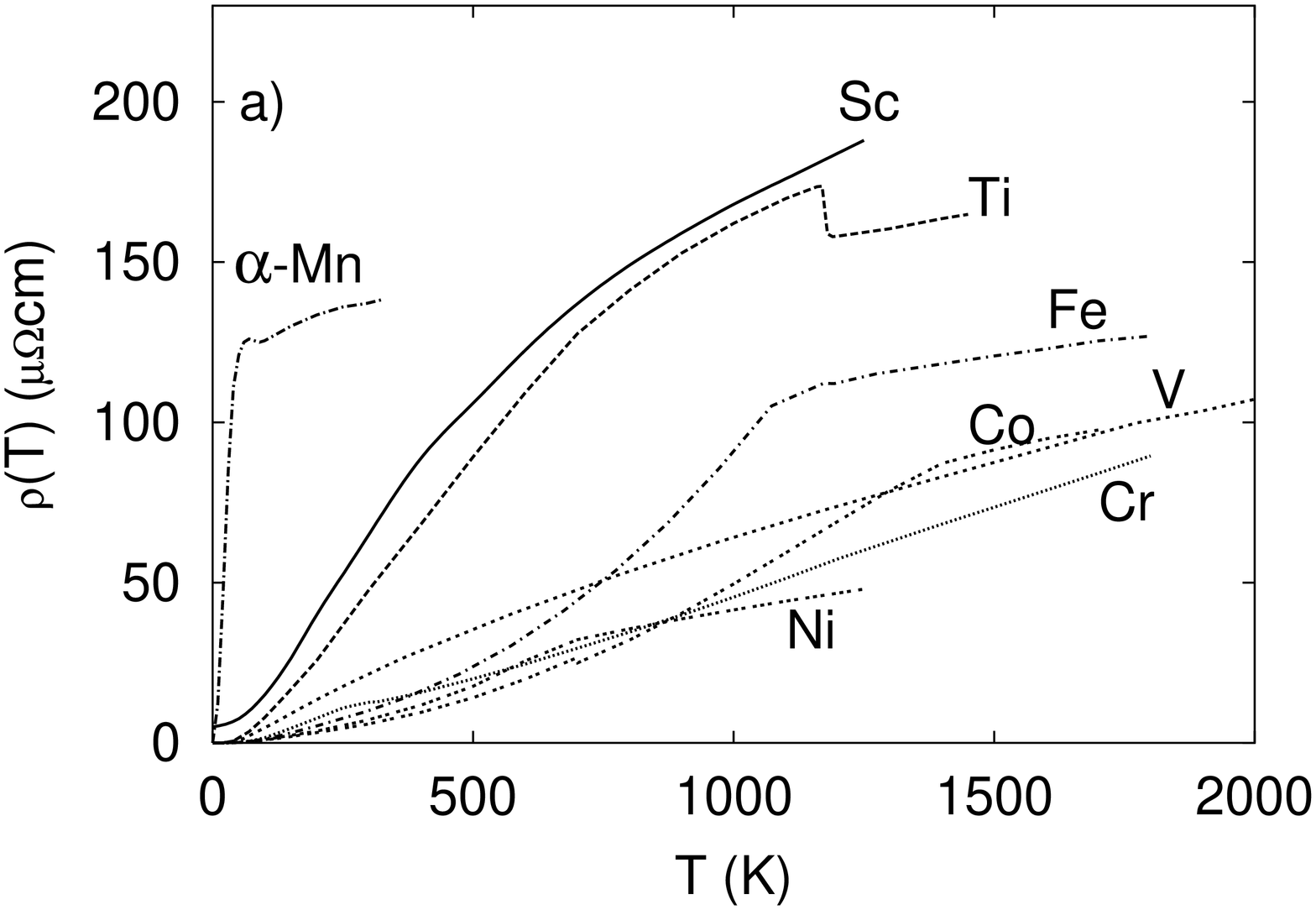}} }}
\centerline{
\rotatebox{-90}{\resizebox{1.9in}{!}{\includegraphics{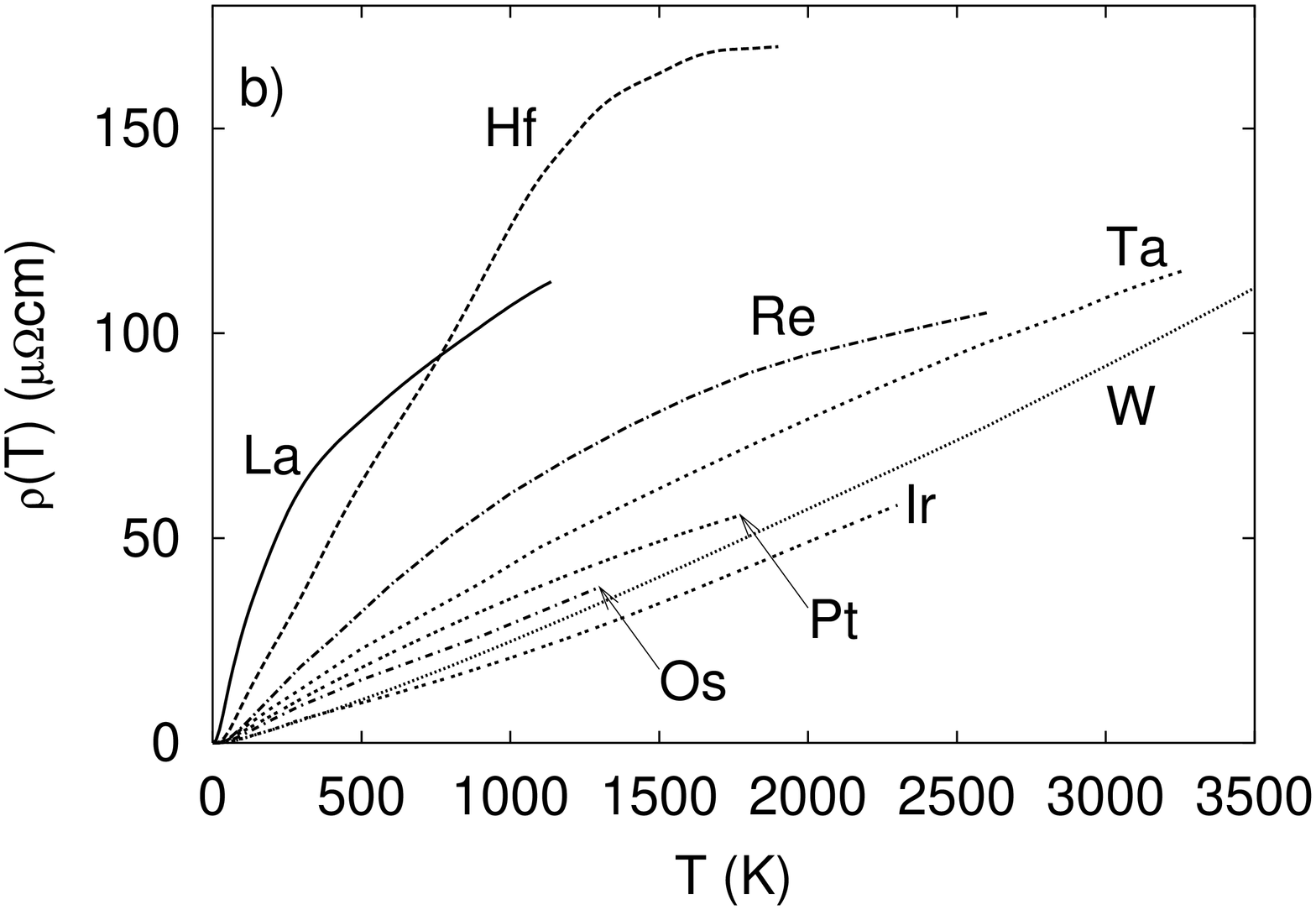}} }}
\caption[]{\label{fig:5}Resistivity of 3d (a) and 5d (b) transition metals
(Bass, 1982).}  
\end{figure}

As discussed in the introduction, resistivity saturation was
found for A15 compounds (Woodard and Cody, 1964; Marchenko, 1973; 
Fisk and Webb, 1976; Wiesmann {\it et al.}, 1977; Gurvitch {\it et al.}, 
1978). Another class of systems showing resistivity saturation 
is the Chevrel phases (Martin {\it et al.}, 1978; Sunandana, 
1979). Early measurements gave very large values for the resistivity, 
probably because of sample problems, but later measurements 
on single crystals gave resistivities comparable to those of A15 
compounds (Kitazawa {\it et al.}, 1981). Fig. \ref{fig:5} shows the 
resistivity of the 3d and 5d transition metals. Several of these 
metals show clear signs of saturation, in particular $\alpha$-Mn. Other
cases are less clear, and, e.g., the resistivity of W even bends slightly 
upwards, although it is already large. As discussed at the end of this 
section, W is nevertheless an example of saturation, if anharmonic 
effects are considered. The above metals can be considered as cases 
where the Ioffe-Regel condition is satisfied.

We next consider the in-plane resistivity of high-$T_c$ cuprates.
The resistivity depends strongly on the doping of the CuO$_2$ plane. 
Information about the 
doping is given by the superconductivity transition temperature $T_c$, 
which is quoted below, when available. As the doping is reduced below
the optimum value, $T_c$ drops and $\rho$ increases. In the case 
of La$_{2-x}$Sr$_x$CuO$_4$  we assume that the doping is given by $x$.
Results for several cuprates were given in Fig. \ref{fig:2}, and we 
give some further examples in Table \ref{table1}. 
These are examples of systems where the resistivity is comparable
to the Ioffe-Regel resistivity already at moderate values of $T$ and 
without showing signs of saturation.

\begin{minipage}{8.6cm}
\begin{table}
\caption{Resistivity $\rho(T)$ (in m$\Omega$cm) of high-$T _c$ 
cuprates. The measurment temperature $T$ and the superconductivity
transition temperature $T_c$ are given in K.\label{table1}}  
\begin{tabular}{lllll}
Compound & $T_c$  & $T$ &$\rho(T)$  & Reference \\
\hline
HgBa$_2$Ca$_0$Cu$_1$O$_{4+x}$ & 94 & 300 & 0.5  &  Daignere 
{\it et al.}, 2001 \\
HgBa$_2$Ca$_1$Cu$_2$O$_{6+x}$ &122 & 300 & 0.3  &  Yan      
{\it et al.}, 1998 \\
HgBa$_2$Ca$_2$Cu$_3$O$_{8+x}$ & 125 & 500 & 0.6  &  Carrington
{\it et al.}, 1994 \\
HgBa$_2$Ca$_3$Cu$_4$O$_{10+x}$ & 130 & 400 & 0.5  &  L\"ohle   
{\it et al.}, 1996 \\
Tl$_2$Ba$_2$CuO$_{6+y}$        &  80 & 300 & 1.3  &  Kubo       
{\it et al.}, 1991\\
Tl$_2$Ba$_2$CuO$_{6+y}$        &  80 & 270 & 0.6  &  Duan      
{\it et al.}, 1991\\
TlSr$_2$CaCu$_2$O$_{7-y}$      &  65 & 300 & 0.5  &  Kubo      
{\it et al.}, 1991\\
Bi$_2$Sr$_2$CaCu$_2$O$_{8+y}$  & 76  & 300 & 1.2  &  Chen      
{\it et al.}, 1998\\
\end{tabular}
\end{table}
\end{minipage}

Bi$_2$Sr$_2$Ca$_{1-x}$Y$_x$Cu$_2$O$_{8+y}$ is of particular interest,
since Wang {\it et al.} (1996ab) found saturation for the most 
underdoped samples, e.g., $x=0.11$ ($T_c=30$ K) as shown in Fig. 
\ref{fig:2}. Similar results have also been obtained by Forro 
{\it et al.} (2002). Other measurements only showed weak signs of 
saturation (Mandrus {\it et al.}, 1992; Ruan {\it et al.}, 2001). 
In the measurements of Ruan {\it et al.} (2001), the reason may have 
been that their $x=0.11$ sample had a higher $T_c=56$ K than that of 
Wang {\it et al.} (1996ab) ($T_c=30$ K), suggesting that the former 
sample was less underdoped. We conclude that the Ioffe-Regel condition
is violated for high-$T_c$ superconductors but that some systems with
very large resistivities may show signs of saturation at much larger 
values than the Ioffe-Regel resistivity.

\begin{figure}[t]
\centerline{
\rotatebox{-90}{\resizebox{2.0in}{!}{\includegraphics{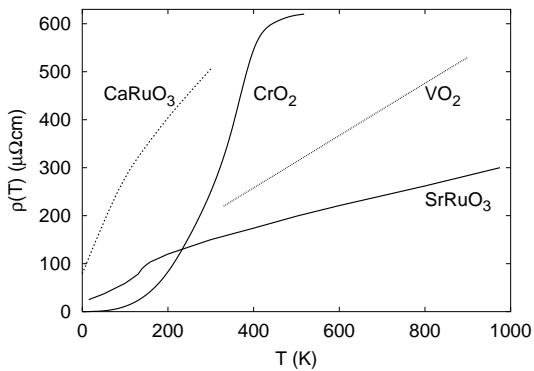}} }}
\caption[]{\label{fig:fig6}Resistivity of CaRuO$_3$ (Klein {\it et al.},
1999ab), CrO$_2$ (Rodbell {\it et al.}, 1966), VO$_2$ (Allen {\it et al.}, 
1993) and SrRuO$_3$ (Allen {\it et al.}, 1996).  }
\end{figure}

Apart from the high-$T_c$ cuprates, a number of transition metal compounds 
have been found which violate the Ioffe-Regel condition. Fig. \ref{fig:3}
and Fig. \ref{fig:fig6} shows some examples. Some of these metals 
show signs of saturation (La$_4$Ru$_6$O$_{19}$, CrO$_2$ and perhaps 
CaRuO$_3$), but at higher resistivities than the Ioffe-Regel resistivity,
while other systems show no sign of saturation. Several of 
theses metals are believed to be non-Fermi liquids. The data in Figs. 
\ref{fig:3} and \ref{fig:fig6} were obtained from single crystals, except 
for Rb$_3$C$_{60}$ and CaRuO$_3$. This could be a reason why the
resistivity in Fig. \ref{fig:fig6} is much larger for CaRuO$_3$ 
than for the related SrRuO$_3$. Large resistivities are also found
in manganites (Salamon and Jaime, 2001). These systems raise interesting
problems in terms of the colossal magnetoresistance, possibly 
important polaron effects and many phase transitions, but they are 
outside the scope of this paper.   

The resistivity of A$_3$C$_{60}$ (A= K, Rb) shows a quite substantial 
spread between different measurements (Hebard {\it et al.}, 1993;
Hou {\it et al.}, 1993; Palstra {\it et al.}, 1994; Degiorgi {\it
et al}, 1994). These have been performed for thin films and doped 
single crystals using direct and optical methods. In particular
the thin films may have substantial defects (e.g., grain boundaries)
which would increase the resistivity. The optical measurements
should, however, be much less sensitive to this. Indeed, optical 
measurements for doped single crystals (Degiorgi {\it et al.}, 1994) 
gave resistivities that are about a factor of two smaller than those
obtained from direct measurements for thin films (Palstra {\it et 
al.}, 1994), but comparable to (somewhat larger than) direct measurements
for doped single crystals (Hou {\it at al.}, 1993).  This suggests 
that the thin film results in Fig. \ref{fig:3} could overestimate 
the resistivity of Rb$_3$C$_{60}$ by a factor of two or somewhat more. 
This would not change the qualitative conclusion  
that the Ioffe-Regel condition is violated for A$_3$C$_{60}$.

Experiments for Rb$_3$C$_{60}$ show no signs of saturation up to
about 500 K (Hebard {\it et al.}, 1993). It is difficult to reach
higher values of $T$, due to the possibility of thermally driven 
rearrangements of the alkali atoms. Hou {\it et al.} (1995), using
a pulsed heating technique to reach 800 K, found a small change 
in the slope at about $T=500$ K. This was interpreted as a sign of 
saturation. Using the parallel resistor formula (Wiesmann {\it et al.}, 
1977), they deduced the saturation resistivity $6\pm 3$ m$\Omega$cm,  
corresponding to the mean free path $l=1\pm0.5$ \AA. Thus there is 
no saturation at $l \sim d$, where $d=10$ \AA \ is the separation of
two C$_{60}$ molecules (see Appendix A), but it is hard to judge 
if there is saturation at some larger resistivity. 

The Boltzmann equation predicts $\rho(T)\sim T$ for large $T$
only if various parameters, such as the electron-phonon interaction, 
are independent of $T$. This assumption is sometimes strongly 
violated. For instance, anharmonic effects reduce the phonon 
frequencies of W by about 30 $\%$ at the melting 
point (Grimvall, 2001; Grimvall {\it et al.}, 1987; Guillermet and 
Grimvall, 1991). According to the Boltzmann equation, this 
should increase $\rho(T)$  by about a factor of two, 
due to an increase in the electron-phonon coupling, giving a growth
of $\rho(T)$ which is much faster than linear. Experimentally,
however, the growth is only slightly faster than linear 
(Fig. \ref{fig:5}b). This suggests that saturation in W is masked 
by the reduction of the phonon frequencies (Grimvall, 2001).

The resistivity is usually measured at constant pressure.
As $T$ is increased, the crystal expands and various parameters
are changed. Sundqvist and Andersson (1990) converted the resistivity
of YBa$_2$Cu$_3$O$_{7-y}$ to constant volume data, which is the more
relevant quantity for theoretical discussions. They found
that this lowered the $T=500$ K resistivity by about 32 $\%$.
Although this conversion involves many uncertainties (Sundqvist
and Andersson, 1990), it was concluded that the pronounced 
linearity in the constant pressure data is lost. Actually, the
corrected data show signs of saturation. A similar correction
of the data for Rb$_3$C$_{60}$ (Vareka and Zettl,
1994) changed the approximately  quadratic $T$ dependence found 
for constant pressure (see Fig. \ref{fig:3}) to an approximately 
linear dependence for constant volume. 

\section{Early theoretical work}\label{sec:c}
Resistivity saturation was studied theoretically very intensively
from the middle of the 1970's to the early 1980's. In this section
we review some of this work as well as some later work. We focus
on work related to the electron-phonon scattering, 
while work based on electron-electron scattering (Jarrell and Pruschke, 
1994; Lange and Kotliar, 1999; Parcollet and Georges, 1999; Merino 
and McKenzie, 2000; Gunnarsson and Han, 2000) fall outside the scope 
of this review. 

Much of the early work has been reviewed by Allen (1980ab).     
Initially, explanations were proposed in terms of strong variations 
of the electronic structure on the energy scale of $k_BT$ or 
unusual anharmonic effects. These effects appear, however, to be 
too small to explain experiment, and as ever more examples of 
resistivity saturation were discovered, such theories had to be 
abandoned (Allen, 1980ab). 

Wiesmann {\it et al.} (1977) found empirically that the resistivity 
of systems such as Nb$_3$Sb can be rather accurately described by 
a parallel resistor formula
\begin{equation}\label{eq:rb1}
{1\over \rho(T)}= {1\over \rho_{\rm sat}}+{1\over \rho_{\rm ideal}(T)},
\end{equation}
where $\rho_{\rm sat}$ is a saturation resistivity and $\rho_{\rm 
ideal}(T)$ is the resistivity of the Boltzmann equation, i.e., linear 
in $T$ for large $T$. 

Mooij (1973) observed that strong disorder can lead to a negative
slope of $\rho(T)$ (Fig. \ref{fig:4}). Jonson and Girvin (1979), 
Girvin and Jonson (1980) and Imry  (1980) argued that this is due 
to an incipient Anderson transition. The disorder is assumed to be 
so large that the system is close to an Anderson transition for $T=0$. 
As $T$ is increased, inelastic scattering (phonon assisted hopping)
leads to a loss of phase information (Lee and Ramakrishnan, 1985)
and the system moves away from the Anderson transition. This was argued 
to lead to a reduction of the resistivity, as observed by Mooij (1973) 
for strongly disordered systems.

Weger and Mott (1985) and Weger (1985) proposed a theory in terms 
dehybridization of the $s$- and $d$-electrons at large values of $T$.
Laughlin (1982) presented a theory where exchange interaction reduced 
the density of states and increased the Fermi velocity in such a way 
that resistivity saturation was obtained.

Cote and Meisel (1978) and Morton {\it et al.} (1978) proposed
that an electron is only scattered by a phonon if the mean free
path of the electron is longer than the phonon wave length. As $T$ 
is increased and $l$ becomes shorter, an increasing number of phonons 
become inefficient scatters. With these assumptions, resistivity 
saturation can be derived. Later calculations treating the electrons 
quantum mechanically, as described before Eq.  (\ref{eq:reb6}), do not 
seem to support the basic assumption in this theory (Calandra and 
Gunnarsson, 2002). 

Several theories have been based on models with one
electronic state per unit cell.  Christoph and Shiller (1984) 
studied the Fr\"ohlich Hamiltonian. Using a truncated equation 
of motion approach, they included some higher order terms
in the electron-phonon interaction. In this way they derived
the parallel resistor formula (Eq. (\ref{eq:rb1})). As pointed
out by them, their $\rho_{\rm sat}\sim 0.84$ m$\Omega$cm is substantially
larger than what is observed for transition metal compounds, but 
it is comparable to what might be expected for a one band model, 
as in Eq. (\ref{eq:reb8}) below ($N_d=1$). Belitz and Schirmacher 
(1983) studied a generalized Fr\"olich model which also included 
impurity scattering. They included higher order terms by using a 
mode coupling technique in a memory function approach. They found 
a rather good agreement with experiment, but a substantially smaller 
$\rho_{\rm sat}$ than might have been expected from Eq. (\ref{eq:reb8}) 
for $N_d=1$. 

Ron {\it et al.} (1986) treated a Kronig-Penney chain, where 
the positions of the $\delta$-functions vibrate.                 
Using the Landauer formula they found saturation in the 
limit of large amplitude vibrations. Their $\rho_{\rm sat}$ depends, 
however, on the strength of the potentials, and it therefore appears 
that the Ioffe-Regel condition is not explained.

The Boltzmann equation only includes intraband transitions. 
Chakraborty and Allen (1979) and Allen and Chakraborty (1981) emphasized 
that resistivity saturation is typically found for systems where 
interband transitions are important. They included these transitions
by generalizing the Boltzmann equation, treating terms of the 
order $(d/l)^0$, while the Boltzmann equation only contains 
terms of the of the order $(d/l)^{-1}$. They demonstrated that 
this leads to a parallel resistor type of formula (Eq. (\ref{eq:rb1})) 
and estimated the saturation resistivity to be of the right order 
of magnitude. Due to the complexity of the equations, it was not 
possible to make explicit calculations and the sign of the $\rho_{\rm 
sat}$ in Eq. (\ref{eq:rb1}) could not be determined. They  
also pointed out that higher order terms in $(d/l)$ may be important. 

\begin{figure}[t]
\centerline{
\rotatebox{-90}{\resizebox{!}{3.5in}{\includegraphics{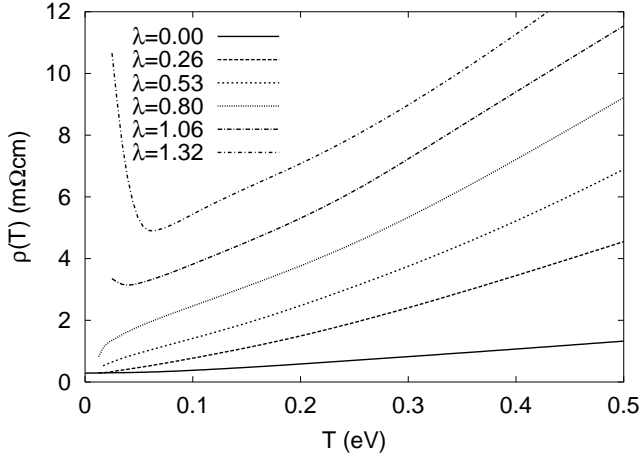}}}}
\caption[]{\label{fig:fig7}Resistivity of the C$_{60}$ model for different
electron-phonon couplings $\lambda$ according to QMC calculations 
(Gunnarsson and Han, 2000; Calandra and Gunnarsson, 2002). The phonon 
frequency is $\omega_{ph}=0.1$ eV.  The figure illustrates the lack of 
saturation. For $\lambda=0.80$ the onset of superconductivity can be seen 
as a sharp downturn in $\rho(T)$ as $T$ is lowered, due to superconducting 
fluctuations. For $\lambda= 1.06$ and 1.32, the resistivity has a negative 
slope for small $T$, indicating an insulating system. }
\end{figure}

Millis {\it et al.} (1999) studied a Holstein like model, where 
the phonons couple to the electron level energies (in the following
called LE coupling).
They found that there is a change of slope in the $\rho(T)$ curve 
for the case of a strong electron-phonon coupling, and associated 
this with resistivity saturation. They observed, however, that 
the resistivity does not saturate and remarked that resistivity 
saturation is a misnomer. Gunnarsson and Han (2000) performed a 
quantum Monte-Carlo (QMC) calculation for a somewhat related model 
of A$_3$C$_{60}$ (A= K, Rb). This model included the partly occupied, 
three-fold degenerate $t_{1u}$ level as well as the LE coupling to a 
five-fold degenerate H$_g$ phonon. We refer to this as the C$_{60}$ model. 
The results, shown in Fig. \ref{fig:fig7}, are somewhat similar
to those of Millis {\it et al.} (1999), but they were considered as an 
example of lack of saturation. It was later shown (Calandra and 
Gunnarsson, 2001) that a more realistic description of saturation
can be obtained by using a coupling to the hopping integrals (HI 
coupling) (see Sec. \ref{sec:e}).

\section{Calculation of resistivity}\label{sec:ca}

The resistivity is usually calculated in the Boltzmann theory.
The electrons are treated semiclassically, and are assumed to
be accelerated by the electric field between scattering processes,
caused by impurities, phonons or other electrons. This can be considered
as the lowest order theory in an expansion in $1/(k_Fl)$ (Kohn and 
Luttinger, 1957). Both because of this and because of its prediction
$\rho(T)\sim T$ for large $T$, it is clear that the Boltzmann equation
cannot be used for describing saturation. Instead, it is 
necessary to treat the electrons quantum-mechanically.
This can be done in the Kubo formalism. For an isotropic system,
this requires the calculation of a current-current correlation 
function (Mahan, 1990)
\begin{equation}\label{eq:1a}
\pi(i\omega)=-{1\over 3N\Omega}\int_0^{\beta}d\tau e^{i\omega \tau}
\langle T_{\tau} {\bf \hat j}(\tau) \cdot {\bf \hat j}(0)\rangle,
\end{equation}
where $N$ is the number of atoms, $\Omega$ is the volume per atom, 
$\beta=1/(k_BT)$, $k_B$ is the Boltzmann constant, $T_{\tau}$ is a 
time-ordering operator and ${\bf j}$ is the current operator. 
$\pi(i\omega)$ is analytically continued to real frequencies,           
giving $\pi_{\rm ret} (\omega)$. The optical conductivity is then 
\begin{equation}\label{eq:1b}
\sigma(\omega)=-{{\rm Im}\pi_{\rm ret}(\omega)
\over \omega},
\end{equation} 
and the resistivity is $1/\sigma(0)$.

$\pi(i\omega)$ can be calculated for imaginary times by using  
a determinantal quantum Monte-Carlo (QMC) method (Blankenbecler 
{\it et al.}, 1981). It is then analytically continued
to real frequencies by using a maximum entropy method (Jarrell and 
Gubernatis, 1996). This approach is very useful for establishing 
the saturation properties of a given model. 

To interpret the results, 
a simpler method is used, where the phonons are treated semiclassically 
(Calandra and Gunnarsson, 2001; 2002). A large supercell with $N$ atoms 
and periodic boundary conditions is considered. Each 
phonon coordinate is given a random displacement according to a 
Gaussian distribution with the width determined by $T$. This set 
of displaced coordinates (snap shot) defines a new Hamiltonian.
While the original Hamiltonian represents a complicated many-body 
problem due to the electron-phonon interaction, the dynamics of the 
phonons have been eliminated in the new Hamiltonian. Since we assume 
no electron-electron interaction, this is then a one-particle 
Hamiltonian for a disordered system. The corresponding eigenstates 
$|l\rangle$ and eigenvalues $\varepsilon_l$ are found. The Kubo-Greenwood
formula (Greenwood, 1958) is then used         
\begin{equation}\label{eq:reb6}
\sigma_{xx}(\omega)={2\pi\over N\Omega \omega}\sum_{ll^{'}}|\langle l|
\hat j_x| l^{'}\rangle |^2(f_l-f_{l^{'}})\delta(\hbar \omega -
\varepsilon_{l^{'}} +\varepsilon_l),
\end{equation}
where $\hat j_x$ is the current operator in the $x$ direction, and 
$f_l$ is the Fermi function for the energy $\varepsilon_l$.  
It is essential that in this approach only the phonons are treated
semiclassically and that the electrons are treated 
quantum-mechanically. This is in contrast to the semiclassical 
treatment of the electrons in the Boltzmann theory, which is unable 
to describe saturation. 

In the following sections we discuss a new theory for describing
different classes of materials with different saturation
properties (Gunnarsson and Han, 2000; Calandra and Gunnarsson, 
2001; 2002; 2003).

\begin{figure}[t]
\rotatebox{-90}{\resizebox{2.0in}{!}{\includegraphics{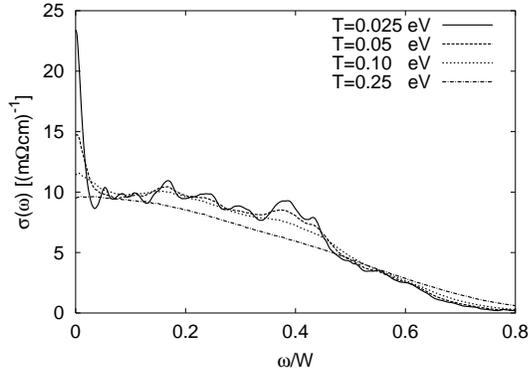}}} 
\caption[]{\label{fig:fig8}
$\sigma(\omega)$ as a function of $\omega$ for the Nb$_3$Sb model in the 
semiclassical treatment of the phonons. The frequency has been scaled  
by the $T=0$ band width $W$. The figure illustrates how the Drude peak 
is lost as $T$ is increased.}
\end{figure}

\section{\lowercase{f}-sum rule}\label{sec:d}

\begin{figure}[t]
\centerline{
\rotatebox{-90}{\resizebox{!}{2.7in}{\includegraphics{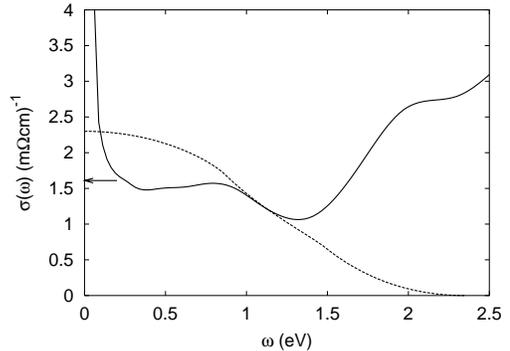}}}}
\caption[]{\label{fig:fig9}$\sigma(\omega)$ for CrO$_2$ according
to experiment (full line) at $T=300$ K (Singley {\it et al.}, 1999) and 
the large $T$ model in Fig. \ref{fig:fig10} (dotted line), 
adjusting the area to the sum rule (\ref{eq:rea1}). The experimental 
intensity above $\omega\sim 1.5$ eV is due to interband transitions. 
The arrow shows the conductivity corresponding to the saturation
resistivity in Fig. \ref{fig:fig6}. It suggests that for large $T$, the 
Drude peak in the experimental $\sigma(\omega)$ disappears.
}
\end{figure}

For small $T$, the optical conductivity $\sigma(\omega)$ typically 
has a Drude peak, due to intraband transitions between quasiparticles 
and centered at $\omega=0$, and a remaining incoherent part, as 
illustrated in Fig. \ref{fig:fig8} for Nb$_3$Sb.  As $T$ is increased, 
the height of the Drude peak is reduced, and for large 
$T$ it disappears. This is also illustrated in Fig. \ref{fig:fig9}, 
which shows the experimental optical conductivity of CrO$_2$ at $T=300$ K. 
The arrow marks the conductivity corresponding to the saturation 
resistivity (see Fig. \ref{fig:fig6}). This conductivity is similar 
to the magnitude of the incoherent contribution (at 300 K). Again,      
this suggests that as $T$ is   increased, the Drude peak disappears 
without the incoherent part changing very much.  In the following, we 
therefore assume that $T$ is so large that the Drude peak has disappeared
(Calandra and Gunnarsson, 2002). 

The Boltzmann theory starts from the periodic system and treats the 
scattering mechanisms as small perturbations. The theory focuses on 
intraband transitions and the Drude peak. Here we start from the    
the opposite limit, where the scattering is so strong that the Drude 
peak disappears.  

To estimate the incoherent contribution to $\sigma(\omega)$, we use 
the f-sum rule. We focus on tight-binding models with one type 
of electrons, e.g., $d$ electrons in a transition metal compound. 
Then there are no on-site matrix elements of the current operator. 
The off-site matrix elements can be related to hopping matrix 
elements, using charge and current conservation.  For an isotropic 
three-dimensional system, the tight-binding version of the f-sum rule 
then takes the form (Maldague, 1977; Calandra and Gunnarsson, 2002)
\begin{equation}\label{eq:rea1}
\int_0^{\infty} \sigma(\omega)d \omega=
{\pi\over 6}{d^2e^2 \over N\Omega\hbar^2}|E_K|,
\end{equation}
where $E_K$ is the hopping (kinetic) energy, $N$ is the number 
of atoms and $\Omega$ is the volume per atom. 
For a system with just nearest neighbor hopping and fixed atoms, 
$d$ is the separation of the sites. If the atoms vibrate    
or if there is hopping between more distant neighbors, $d$ is a weighted 
average of neighboring distances (Calandra and Gunnarsson, 2002).

\begin{figure}[t]
\centerline{
\resizebox{2.0in}{!}{\includegraphics{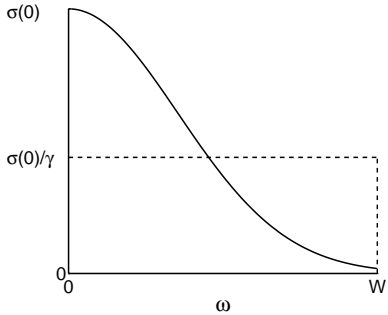}} }
\caption[]{\label{fig:fig10}
Schematic picture of $\sigma(\omega)$. The average of $\sigma(\omega)$   
over the band width is given by $\sigma(0)/\gamma$, where $\gamma\sim 2$.
}
\end{figure}

For models with only one type of electrons, e.g, $d$ electrons,     
$\sigma(\omega)$ at large $T$ looks schematically as in Fig. 
\ref{fig:fig10}. This can be compared with the calculation for Nb$_3$Sb 
in Fig. \ref{fig:fig8} or the expected large $T$ contribution from the 
$t_{2g}$ band of CrO$_2$ in Fig.  \ref{fig:fig9} (Drude peak gone),
excluding the interband transitions for $\hbar\omega > 1.5$ eV. For 
$\hbar\omega$ larger than the band width $W$, $\sigma(\omega)\approx 0$. 
If $\sigma(\omega)$ furthermore lacks large structures around $\omega=0$, 
we can approximate 
\begin{equation}\label{eq:rea2}
\sigma(\omega)= \cases{ \sigma(0)/\gamma,&if $\hbar|\omega|\le W$;\cr
0, &  otherwise.\cr}
\end{equation}
Then $\int \sigma(\omega)\hbar d\omega=\sigma(0)W/\gamma$, 
and the f-sum rule (\ref{eq:rea1}) requires that $\sigma(0)$
is given by the right hand side of Eq. (\ref{eq:rea1})
multiplied by $\gamma\hbar/W$ (Calandra and Gunnarsson, 2002; 2003), i.e., 
\begin{equation}\label{eq:rea3}
\sigma(0)={\gamma \hbar\over W}\int_0^{\infty} 
\sigma(\omega)d \omega=   {\pi\gamma\over 6W}{d^2e^2 
\over N\Omega\hbar}|E_K|,
\end{equation}
where $\gamma\sim 2$ depends on the shape of $\sigma(\omega)$. 
This provides an approximate lower limit to the conductivity 
(upper limit to the resistivity), which may be reached 
if $T$ is so large that the Drude peak is negligible. The conductivity 
could be appreciably smaller only if there is a substantial amount of 
weight for $\omega>W$ or if $\sigma(\omega)$ is anomalously small for  
$\omega=0$. The former should not be important, since 
$\hbar\omega>W$ requires multiple electron-hole pair excitations. Since 
the current operator is a one-particle operator, these excitations  
should have a small weight. The latter could happen in the case of an 
incipient Anderson transition, but it should normally not occur at large 
$T$, where inelastic scattering destroys the phase information required 
for an Anderson localization (Lee and Ramakrishnan, 1985). The (lower) limit 
(\ref{eq:rea3}) to the conductivity is $T$ dependent, due to the 
$T$ dependence of $E_K$ and $W$. This $T$ dependence is often, but not 
always, rather weak.  Below we use Eq. (\ref{eq:rea3}) to discuss
different classes of metals.

\section{Weakly correlated broad band systems}\label{sec:e}

We first focus on metals with a large band width $W$. We therefore
assume that i) $W$ is so large that the electrons can be treated 
as noninteracting and ii) $T\ll W$. 
In particular, we focus on  Nb$_3$Sb, showing a pronounced saturation, 
and Nb, showing a much weaker saturation (see
Fig. \ref{fig:1}). We use a model with an $N_d$-fold degenerate orbital 
on each site. The orbitals on different atoms are connected by hopping 
matrix elements, $t_{im,jm^{'}}$, where $i$ and $j$ label the atoms and 
$m, m^{'} = 1, .., N_d$ the orbital. Then the electronic Hamiltonian is 
\begin{equation}\label{eq:reb1}
H^{\rm el}=\sum_{i\ne j,mm^{'}\sigma}t_{im,jm^{'}} 
c^{\dagger}_{im\sigma}c_{jm^{'}\sigma},
\end{equation}
Three Einstein phonons are introduced for each atom, describing its 
displacement in the three coordinate directions. The phonons couple 
to the hopping integrals, since these depend on the atomic positions. 
We refer to this as HI coupling and the model is referred to as the 
transition metal (TM) model. The model for Nb has $4d$ orbitals 
($N_d=5$) on a bcc lattice. For Nb$_3$Sb we use the Nb$_3^{\ast}$ model, 
neglecting the Sb atoms (Pickett {\it et al.}, 1979), and putting $4d$ 
orbitals on the Nb sites. The models are therefore identical except 
for the lattice structure. For strongly ionic systems, the coupling 
to the level energies (LE) may also be important.

\begin{figure}[t]
\centerline{
\resizebox{2.5in}{!}{\includegraphics{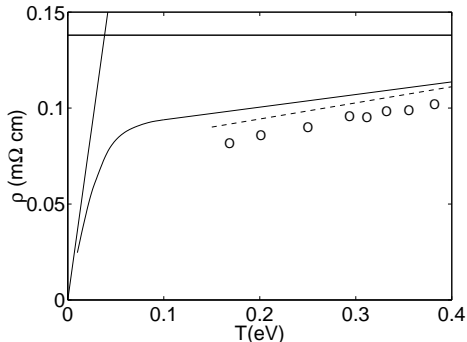}} }
\caption[]{\label{fig:fig11}Resistivity for the Nb$_3^{\ast}$ model of
Nb$_3$Sb, comparing the semiclassical $\lbrack$ broken ($N=36$ atoms) 
and full ($N=648$) curves$\rbrack$ and QMC (circles, $N=36$) treatment 
of the phonons. The figure also shows the 
small (Eq. (\ref{eq:reb9})) and large (Eq. (\ref{eq:reb8})) $T$         
results. It illustrates that the resistivity of the TM model 
saturates at large $T$ (after Calandra and Gunnarsson, 2001).}
\end{figure}

Fig. \ref{fig:fig11} shows the QMC (circles) results for the resistivity
of Nb$_3^{\ast}$ model for a supercell with $N=36$ atoms. These results 
extrapolate to a large resistivity 
at $T=0$. On the other hand, the resistivity of the model must be zero at 
$T=0$. It then follows that there must be a large change of the slope 
of $\rho(T)$ as $T$ is increased, i.e., resistivity saturation. 
This model can therefore be used for analyzing saturation. Fig. 
\ref{fig:fig11} also shows that the semiclassical treatment of the
phonons is quite accurate for the TM model, and that it therefore
can be used for interpreting the results.     

To use f-sum rule of Sec. \ref{sec:d}, we calculate the hopping energy 
$E_K$. Since we have assumed that the electrons are noninteracting 
and that $T\ll W$, we obtain 
\begin{equation}\label{eq:reb7}
E_K=2N_d\int_{-W/2}^{E_F}\varepsilon
N(\varepsilon)d\varepsilon\equiv -2\alpha N_d W N,
\end{equation}
where $E_F$ is the Fermi energy, $N(\varepsilon)$ is the density of 
states per atom, orbital and spin, $N_d$ is the orbital degeneracy,
and $\alpha$ depends on the shape 
of $N(\varepsilon)$ and the filling. Typically $\alpha\sim 0.1$ for 
a metal close to half-filling.  The dependence on filling is moderate
around half-filling (Calandra and Gunnarsson, 2002), but $\alpha=0$ 
for an empty or full band.  Inserting Eq. (\ref{eq:reb7}) in
Eq. (\ref{eq:rea3}) gives an approximate lower limit to the 
conductivity (Calandra and Gunnarsson, 2002; 2003)
\begin{equation}\label{eq:reb8}
\sigma_{\rm sat}(0)={\pi \alpha\gamma\over 3} 
{d^3\over \Omega}{N_de^2\over \hbar d}.
\end{equation}
Here $\pi\alpha\gamma/3 \sim 0.2 $ depends on the details of the electronic 
structure, and $d^3/\Omega\sim 1$ depends on the lattice structure. The 
result is independent of $W$, since the $W$ in Eq. (\ref{eq:rea3}) cancels 
that in Eq. (\ref{eq:reb7}). The essential material parameters, $d$
and $N_d$ appear in $N_de^2/(\hbar d)$, which has the dimension conductivity.
The corresponding apparent mean free path is (Calandra and Gunnarsson,
2001)
\begin{equation}\label{eq:reb8a}
l=cN_d^{1\over 3}d,
\end{equation}
where $c\sim 0.6-0.8$, which is an approximate lower limit to $l$. It 
then follows that $l \gtrsim d$. This provides a quantum-mechanical 
derivation of the Ioffe-Regel condition, assuming that the electrons 
can be treated as noninteracting and that $T\ll W$. For a transition metal 
compound, with $N_d=5$ and $d\sim 3$ \AA, this leads to a resistivity of 
the order of 0.1-0.2 m$\Omega$cm, in agreement with the experiments.

We next consider the small $T$ behavior of $\rho(T)$. In this limit 
the Boltzmann theory can be used, since $l\gg d$. For $T$ larger than 
some fraction of $\omega_{ph}$ (Grimvall, 1981) 
\begin{equation}\label{eq:reb9}
\rho(T)= 8\pi^2{\lambda T k_B\over \hbar \Omega_{pl}^2},
\end{equation}
where $\lambda$ is the dimensionless electron-phonon coupling constant 
and $\Omega_{pl}$ is the plasma frequency
\begin{equation}\label{eq:reb10}
(\hbar\Omega_{pl})^2={e^2\over 3\pi^2}\sum_{\nu} \int_{Bz}d^3k \lbrack 
{\partial \varepsilon_{\nu}({\bf k}) \over \partial {\bf k}}\rbrack^2\delta
(\varepsilon_{\nu}({\bf k})-E_F).
\end{equation}
Here $\varepsilon_{\nu}({\bf k})$ is the energy of a state with
the band index $\nu$ and the wave vector ${\bf k}$. $\Omega_{pl}$ 
depends on the average Fermi velocity.

The straight line corresponding to Eq. (\ref{eq:reb9}) is shown 
in Fig. \ref{fig:fig11}. It shows how the conductivity is reduced 
and the resistivity is increased due to the reduction of the Drude 
peak height as $T$ is increased. The horizontal line in Fig. 
\ref{fig:fig11} corresponds to the large $T$ limit (Eq. (\ref{eq:reb8})) 
of the conductivity. When this horizontal line is crossed by the 
steep line (Eq. (\ref{eq:reb9})), the Drude peak has been so strongly 
reduced that it disappears under the incoherent part, and is spread 
out over a large energy range. As discussed above, due to the f-sum rule, 
the conductivity cannot normally be much further reduced at this point. 

Saturation requires that the two lines (Eqs. (\ref{eq:reb8}, 
\ref{eq:reb9})) cross in the available temperature range. For most 
metals, the line in Eq. (\ref{eq:reb9}) has such a small slope that 
the crossing would happen far above the melting point.

The model of Nb shows a much weaker saturation than the Nb$_3^{\ast}$ 
model of Nb$_3$Sb (Calandra and Gunnarsson, 2001), as is also found 
experimentally (see Fig. \ref{fig:1}). This is due to a much larger 
$\Omega_{pl}$ for Nb (9.5 eV in the model) than for Nb$_3$Sb (3.4 eV). 
The slope of the line in Eq. (\ref{eq:reb9}) is therefore a factor of 
five larger for Nb$_3$Sb than for Nb. This leads to the much more
pronounced saturation for Nb$_3$Sb. The difference in $\Omega_{pl}$
is due to the large unit cell of Nb$_3^{\ast}$, which leads to            
many bands and many forbidden crossings. This results in quite flat 
bands and small electron velocities and a small $\Omega_{pl}$
(Eq. (\ref{eq:reb10})). In this context it is interesting to 
note that $\alpha$-Mn has the most pronounced saturation among
the transition metals (see Fig. \ref{fig:5}a). The unit cell of 
$\alpha$-Mn has 58 atoms, suggesting a very small $\Omega_{pl}$ 
and a very steep initial slope of $\rho(T)$. Chakraborty and Allen 
(1979) and Allen and Chakraborty (1981) observed that saturation 
usually happens for systems with important interband transitions. 
Such systems typically have large unit cells and small $\Omega_{pl}$. 
Their observation can then be understood in terms of the arguments above. 

For a one-band model with one $s$-state per unit cell, 
$\rho_{\rm sat}$ is large ($N_d=1$ in Eq. (\ref{eq:reb8}) instead of 
$N_d=5$ for the TM model) and 
$\lambda/\Omega_{pl}$ is typically small. The crossing of the two
lines in Eqs. (\ref{eq:reb8}, \ref{eq:reb9}) then happens for such  
large vibration amplitudes that a calculation is not very meaningful.
The main reason is that for an $s$-band model, the matrix elements 
do not change sign as the atoms vibrate, and very large vibration 
amplitudes are needed to make the scattering so strong that the Drude 
peak essentially disappears. It is therefore interesting 
to study a one-band model where the state has many strongly directed 
lobes. Even moderate atomic vibrations then change the signs of 
the matrix elements. As a model we use one of the three $t_{1u}$ 
orbitals of a C$_{60}$ molecule (hypothetical case of a strong 
crystal-field splitting). This orbital has $l=5$ and correspondingly 
many nodes. The resulting resistivity saturation is moderate, with 
the slope at large $T$ being reduced by about a factor of three to four 
in the semiclassical treatment of the phonons. This illustrates that 
interband transitions are not indispensable for resistivity saturation.  

The optical conductivity in Figs. \ref{fig:fig8} and \ref{fig:fig9} 
can be viewed as a Drude peak riding on top of a broad 
structure of incoherent contributions. According to the f-sum rule and 
the arguments above, the height and shape of the latter part does not 
change much for values of $T$ discussed here.  
We then approximately have   
\begin{equation}\label{eq:d17}
\sigma(\omega=0,T)=\sigma_{\rm Drude}(\omega=0,T)+\sigma_{\rm sat},
\end{equation}
where $\sigma_{\rm Drude}(\omega=0,T)$ is the inverse of the resistivity
in Eq. (\ref{eq:reb9}) and $\sigma_{\rm sat}$ is the height of the 
incoherent contribution (Eq. (\ref{eq:reb8})). This is the parallel resistor 
formula (Wiesmann {\it et al.}, 1977). 

\section{Strongly correlated systems. High-$T_{\lowercase{c}}$ cuprates}\label{sec:f}

In the derivation of the Ioffe-Regel condition, we assumed i) noninteracting
electrons. We now focus on the high-$T_c$ cuprates, where this assumption
is invalid. The transport properties in the CuO$_2$ plane are mainly 
determined by the antibonding band of Cu $x^2-y^2$ and O $2p$ character. 
It is then natural to consider one ($x^2-y^2$) orbital per site and the 
orbital degeneracy $N_d=1$. These orbitals are put on a two-dimensional 
square lattice, describing the CuO$_2$ plane. The Coulomb 
interaction is described by a Hubbard $U$ acting between two electrons 
on the same site. $U$ is assumed to be so large that states with double 
occupancy on a site can be projected out, leading to the $t-J$ model 
(Zhang and Rice, 1988).  

\begin{figure}[t]
\centerline{
\rotatebox{-90}{\resizebox{!}{3.0in}{\includegraphics{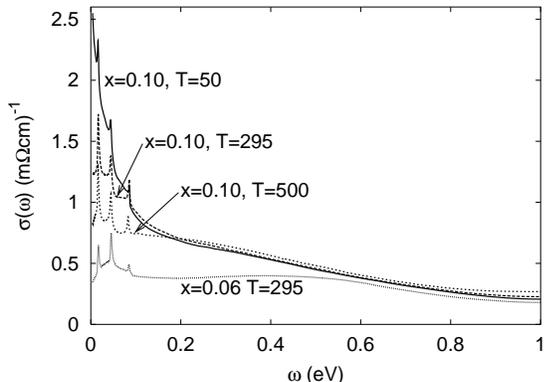}}}}
\caption[]{\label{fig:fig12} 
In-plane optical conductivity of La$_{2-x}$Sr$_x$CuO$_4$ as a function of 
$\omega$ for $x=0.06$ and $x=0.10$ (Takenaka {\it et al.}, 2002). The 
figure illustrates how the peak at $\omega=0$ gradually goes away as $T$ 
is increased and that the intraband contribution to $\sigma(\omega)$ 
is reduced as $x$ is reduced. The sharp structures at small $\omega$ are 
due to phonons. Interband transitions become important for $\hbar \omega
\gtrsim 1$ eV. }
\end{figure}

To use the f-sum rule, we first make a simple estimate of the hopping 
energy. Due to the doping with $x$ holes per site, a given site is 
occupied by a hole or an electron with the probabilities $x$ and $(1-x)$, 
respectively. In the $t-J$ model, a hole can hop to a neighboring site 
only if the site is occupied by an electron. This suggests a hopping 
probability of the order of $x(1-x)$. Multiplying by the number of 
nearest neighbors (four) and the hopping integral $t$ gives a simple 
estimate of the hopping energy (Calandra and Gunnarsson, 2003)
\begin{equation}\label{eq:rec2}
 E_K =-4tx(1-x)N,
\end{equation}
where $N$ is the number of sites. An exact diagonalization calculation 
for the $t-J$ model, gives a similar dependence on $x$ but a
somewhat smaller prefactor of about 3.4 (Calandra and Gunnarsson, 2003). 

Typically, $x \ll 1$. From Eq. (\ref{eq:rec2}) it then follows that 
the right hand side of the 
f-sum rule (Eq. (\ref{eq:rea1})) should be approximately proportional 
to $x$, as is also found experimentally (Yamada {\it et al.}, 1998). 
The constant of proportionality, extracted from  Uchida {\it et al.} 
(1991), agrees with the calculation to within about 20-30 $\%$ (Calandra 
and Gunnarsson, 2003). 

Fig. \ref{fig:fig12} shows $\sigma(\omega)$ of La$_{2-x}$Sr$_x$CuO$_4$
for different $T$ and $x$ (Takenaka {\it et al.}, 2002).
The figure illustrates how the peak at $\omega=0$ gradually disappears 
as $T$ is increased and that for large $T$, apart from phonon structures, 
$\sigma(\omega)$ behave roughly as assumed in Fig. \ref{fig:fig10}. 
A comparison of the curves
for $x=0.06$ and $x=0.10$, shows that $\sigma(\omega)$ is much smaller
for $x=0.06$ in the energy range $\omega\lesssim 1$ eV, where the 
intraband contributions dominate. This illustrates the reduction of 
the intraband contribution to $\sigma(\omega)$ when $x$ is reduced, 
as assumed above. Similar results were found for 
La$_{2-x}$(Ca,Sr)$_x$CaCuO$_{6+\delta}$ 
by Wang {\it et al} (2002).

We now analyze the resistivity in terms of the f-sum rule (Eqs. 
(\ref{eq:rea1}, \ref{eq:rea3})).  For conduction in the 
ab-plane of a cuprate, the factor $1/6$ in Eqs. (\ref{eq:rea1},
\ref{eq:rea3}) should be replaced by a factor $1/4$, since the 
the system is quasi-two-dimensional.                            
Insertion of Eq. (\ref{eq:rec2}), but with the prefactor 3.4, 
in Eq. (\ref{eq:rea3}) gives 
\begin{equation}\label{eq:rec3}
\rho_{\rm sat}={0.07 c \over x(1-x)} \approx 
{0.4\over x(1-x)} \hskip0.3cm {\rm m}\Omega{\rm cm},
\end{equation}
where we have used distance $c=6.4$ \AA \ between the CuO$_2$ planes
appropriate for La$_{2-x}$Sr$_x$CuO$_4$. For small $x$, the saturation 
resistivity in Eq. (\ref{eq:rec3}) is very much larger than the Ioffe-Regel 
resistivity, 0.7 m$\Omega$cm. It is also much larger than the 
result in Sec. \ref{sec:e} for weakly correlated transition metal 
compounds (Eq. (\ref{eq:reb8}))
\begin{equation}\label{eq:rec4}
\rho_{\rm sat} \sim {0.14 d\over N_d}\approx {0.4 \over N_d} \hskip0.3cm 
{\rm m}\Omega{\rm cm},
\end{equation}
where $d\sim 3$ is expressed in \AA.  This is partly due to the 
degeneracy being just $N_d=1$ for the $t-J$ model but $N_d=5$ for the 
TM model. Furthermore, the strong correlation drastically reduces the 
hopping energy in the cuprates, which gives the factor $x(1-x)$ in Eq. 
(\ref{eq:rec3}). Finally, the large separation of the CuO$_2$ planes 
also increases $\rho$. 

\begin{figure}[t]
\hskip0.5cm
\rotatebox{-90}{\resizebox{!}{3.5in}{\includegraphics{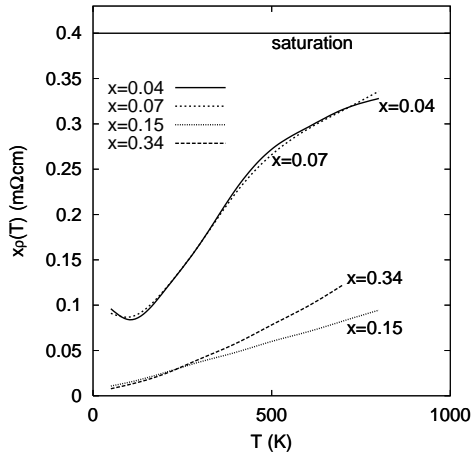}} }
\caption[]{\label{fig:fig13} Resistivity multiplied by the doping
$x$ for La$_{2-x}$Sr$_x$CuO$_4$ (Takagi {\it et al.}, 
1992) for $x=0.04$ (full curve), $x=0.07$ (broken curve), $x=0.15$ 
(dotted curve) and $x=0.34$ (chain curve). The horizontal 
line shows the saturation resistivity (Eq. (\ref{eq:rec3})), 
$x\rho_{\rm sat}$, in the limit of small $x$. The figure illustrates 
that the saturation resistivity is not exceeded by the experimental 
data and that there are signs of saturation for small $x$ roughly 
where saturation is expected.}
\end{figure}

Fig. \ref{fig:fig13} compares the experimental results of Takagi (1992) 
for La$_{2-x}$Sr$_x$CuO$_4$ with the saturation resistivity in Eq. 
(\ref{eq:rec3}). Due to the factor $1/x$ in Eq. (\ref{eq:rec3}), all 
resistivities are multiplied by the doping $x$. For small $x$, 
$\rho_{\rm sat}$ is then a constant. The experimental resistivity is smaller 
than the predicted saturation resistivity. The same conclusion is obtained 
for other high-$T_c$ superconductors (Calandra and Gunnarsson, 2003).
The experimental data do therefore not demonstrate absence of saturation. 
On the contrary, for La$_{2-x}$Sr$_x$CuO$_4$ with $x=0.04$ and $x=0.07$ 
there are signs of saturation where saturation is expected to occur.
It is interesting that the curves $x\rho(T)$ for $x=0.04$ and $x=0.07$ 
fall almost on top of each other, giving a dependence $\rho\sim 1/x$, 
as suggested by the arguments above. For small dopings, saturation has 
also been reported (Wang {\it et al.}, 1996ab) for 
Bi$_2$Sr$_2$Ca$_{1-x}$Y$_x$Cu$_2$O$_{8+y}$ at similar values as for
La$_{2-x}$Sr$_x$CuO$_4$ (see Fig. \ref{fig:2}).

Above we gave an upper limit to the resistivity without 
specifying a scattering mechanism. For $x=0.15$, the resistivity 
in the $t-J$ model is of the right order of magnitude, suggesting
that electron-electron scattering can explain the resistivity, 
while for $x=0.07$ and $x=0.04$ the resistivity is much too small 
compared with experiment(Calandra and Gunnarsson, 2003). This 
suggests that for $x=0.04$ and $x=0.07$, there is an additional 
important scattering mechanism, beyond the electron-electron scattering 
mechanism in the $t-J$ model, or that our small cluster (16 sites) 
cannot describe some important scattering mechanism.  
This mechanism apparently increases the resistivity so rapidly for 
small $T$ that the upper limit (\ref{eq:rec3}) is approached and 
the resistivity shows sign of saturation. The estimate of the upper limit 
(\ref{eq:rec3}) is nevertheless correct, unless the additional 
scattering mechanism appreciably influences the hopping energy. 
The $t-J$ model alone only gives a rather weak saturation and only 
at larger $T$ than seen in Fig. \ref{fig:fig13}.

\section{Violation of the Ioffe-Regel condition in other compounds}
\label{sec:g}

Figs. \ref{fig:3} and \ref{fig:fig6} show that the Ioffe-Regel condition is 
violated for several other compounds. We first discuss CrO$_2$, which 
is a ferromagnet below $T_C=390$ K. According to band structure 
calculations, three bands of mainly Cr t$_{2g}$ spin up character are
occupied by two electrons (Lewis {\it et al.}, 1997; Mazin {\it et al.}, 
1999). We therefore consider a three-fold degenerate band with the width 
$W=2.3$ eV (Lewis {\it et al.}, 1997) and without spin degeneracy. 
Fig. \ref{fig:fig9} compares the experimental $\sigma(\omega)$ at 
$T=300$ K (Singley {\it et al.}, 1999) with the large $T$ theoretical 
model in Fig. \ref{fig:fig10}. The theoretical area, $\hbar^2 \int
\sigma(\omega)d\omega =2.7$ (eV)$^2$ (for units see Appendix A) has 
been obtained from the f-sum rule (Eq. (\ref{eq:rea1})), assuming           
a semi-elliptical band with $W=2.3$ eV. We estimate the experimental 
t$_{2g}$ contribution to the f-sum rule to be 2.2 (eV)$^2$, i.e., somewhat 
smaller. As a result, our estimate of the saturation resistivity
($\rho_{\rm sat}=0.4$ m$\Omega$cm) is somewhat smaller than the
experimental result ($\sim 0.6-0.7$ m$\Omega$cm). The reason is 
probably both that the area of $\sigma(\omega)$ is somewhat 
overestimated and that the theoretical shape of $\sigma(\omega)$ does 
not take into account a strong dip in $N(\varepsilon)$ at $E_F$ 
for CrO$_2$. These considerations suggest that theory gives
a qualitatively correct description of saturation for CrO$_2$. 
The strong increase in $\sigma(\omega)$ for $\omega\sim 1.5$
eV is due to interband transitions, which are neglected in the model.    

This treatment assumed CrO$_2$ to be ferromagnetic. If we had taken
into account that CrO$_2$ is paramagnetic for $T>T_C$, the theoretical
saturation resistivity would have dropped by a factor of two,
due to an increase of $|E_K|$. Experimentally, however, the resistivity 
changes little at $T=T_C$, suggesting that correlation effects reduce 
$|E_K|$ above $T_C$ in a way similar to that of magnetic effects below 
$T_C$. Actually, the Hubbard $U$ of CrO$_2$ has been estimated to 
$U\approx 3$ eV (Korotin {\it et al.}, 1998), comparable to the 
$t_{2g}$ band width, and suggesting appreciable correlation effects. 
There are rather strong magnetic fluctuations in several of the systems 
violating the Ioffe-Regel condition. The arguments above suggest that 
this may reduce $|E_K|$ and increase the saturation resistivity.

Using similar assumptions (correlation reduces $|E_K|$ by a factor
of two in the paramagnetic state) as for CrO$_2$, we predict a similar 
saturation resistivity for VO$_2$ and a somewhat larger value
(0.6 m$\Omega$cm) for CaRuO$_3$ and SrRuO$_3$. As for CrO$_2$, more 
experimental data is needed to decide whether this agrees qualitatively with 
experiment. It is not clear if this framework (strong correlation
effects reducing $|E_K|$)  can be used to understand 
Sr$_2$RuO$_4$, which shows no signs of saturation up to $T=1300$ K, and 
La$_4$Ru$_6$O$_{19}$, which shows signs of saturation, but at a very 
large resistivity.
 
\section{Very large $T$ behavior. C$_{60}$ compounds}\label{sec:h}

To derive the Ioffe-Regel condition, we assumed that ii) $T\ll W$, leading 
to a $T$ independent upper limit $\rho_{\rm sat}$ to the resistivity. 
We now consider $T$ to be so large that the phonons substantially 
change $W$. Such effects are usually not very important, but the C$_{60}$ 
model is an important exception, due to its small band width. The $T$
dependence of $\rho_{\rm sat}$ can then be so strong that the concept
of resistivity saturation becomes meaningless. In the following 
we neglect the electron-electron interaction, although it could
play a substantial role for alkali-doped fullerides. We treat
the phonons in the semiclassical approximation (Sec. \ref{sec:ca}).

For very large $T$, there is a rather trivial $T$ dependence
due to the Fermi temperature, $T_F$, entering in the Fermi-functions
of Eq. (\ref{eq:reb6}). Let us consider the resistivity due to 
static disorder. Expanding the Fermi functions in $1/T$, we
obtain $\sigma(0)\sim 1/T$ and $\rho(T)\sim T$,
although the scattering mechanism itself is $T$-independent. 
A similar dependence also enters for the electron-phonon
scattering, which tends to mask interesting differences between
couplings to the level energies (LE) and hopping integrals (HI). 
We therefore study the case when the Fermi functions are replaced
by $\Theta$-functions in Eq. (\ref{eq:reb6}) ($T_F=0$).

\begin{figure}[t]
\centerline{
\rotatebox{-90}{\resizebox{!}{3.0in}{\includegraphics{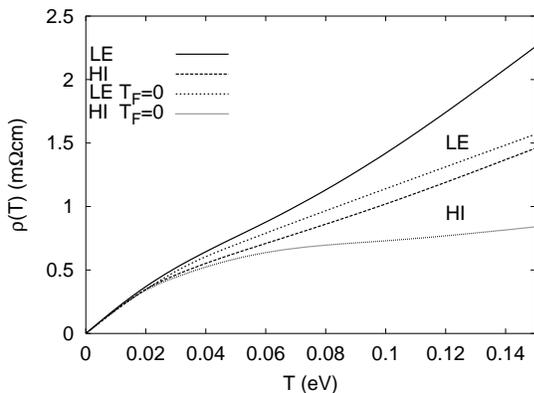}}}}
\caption[]{\label{fig:fig14}Resistivity for a model of ordered C$_{60}$ 
molecules, considering couplings 
to the level energies (full line, LE coupl.) and to the hopping 
integrals (broken line, HI coupling) in the semiclassical treatment 
of the phonons. Results are also shown for $T_F=0$, replacing 
the Fermi functions in Eq. (\ref{eq:reb6}) by $\Theta$-functions. 
The figure shows that there is a large difference between LE and HI 
coupling (after Calandra and Gunnarsson, 2002).}
\end{figure}

We consider the C$_{60}$ model (see Sec. \ref{sec:c}) with either 
LE or HI coupling (Calandra and Gunnarsson, 2002), using the same 
coupling $\lambda$ in both cases (Fig. \ref{fig:fig14}). While the 
resistivity shows no signs of saturation for the LE coupling (full curve),
the model with HI coupling shows a weak saturation (broken curve),
i.e.,  a moderate reduction of the slope for $T\gtrsim 0.03$ eV.
This saturation becomes much more pronounced for $T_F=0$ (dotted curve). 
For the TM model, the slope of $\rho(T)$ is reduced at high $T$ for both 
HI and LE coupling, but the reduction is more pronounced for HI coupling.

To apply the analysis of Sec. \ref{sec:d}, we calculate $W(T)$ and 
$E_K(T)$. $W(T)$ increases with $T$ for both LE and HI coupling. 
In the LE case this happens because the fluctuations in the level 
positions increase and in the HI case because the average of the 
square of the hopping integrals increases. Since the average phonon 
amplitude squared is $\langle x^2\rangle \sim T$, the second moment 
$S_2(T)$ of the density of states goes as $S_2(T)=S_2(0)+aT$, where 
$a$ is some constant. This leads to (Calandra and Gunnarsson, 2002)
\begin{equation}\label{eq:de6}
W(T)=W(0)\sqrt{1+c\lambda {k_B T\over W(T=0)}},
\end{equation}
The $T$ dependence comes in the form $\lambda T/W$, where 
the large prefactor $c=12$ is due to the large ratio $W^2/S_2\sim 12$. 
Using the considerations of Sec. \ref{sec:e}, it follows 
that for the HI coupling $E_K$ varies in a similar way, and the $T$ 
dependences of $E_K$ and $W$ essentially cancel (Eq. (\ref{eq:rea3})). 
This is confirmed by Fig. \ref{fig:fig14} for the C$_{60}$ model 
(lowest HI curve), while the cancellation is less complete for the TM 
model due to a stronger $T$ dependence of $\alpha$ (Eq. (\ref{eq:reb7})) 
and $\gamma$ (Eq. (\ref{eq:rea3})). With LE coupling, on the other hand, 
$E_K$ is reduced as $T$ is increased. As the phonons are excited, the 
level energies on different sites become different and hopping 
becomes more difficult. For the C$_{60}$ model  $|E_K|\sim 1/
\sqrt{1+c\lambda k_BT/W(T=0)}$ (Calandra and Gunnarsson, 2002). 
For the LE coupling, the hopping energy and the band width work 
together, and the upper limit to the resistivity takes the form
\begin{equation}\label{eq:de11}
\rho_{\rm sat}(T)={0.8\over \gamma(T)}(1+c \lambda {k_BT\over W(T=0)})
\hskip0.5cm {\rm m}\Omega{\rm cm}.
\end{equation}

In A$_3$C$_{60}$ (A= K, Rb) there is orientational disorder, i.e.,
the C$_{60}$ molecules more or less randomly take one out of two 
preferred orientations (Stephens {\it et al.}, 1991). As a result, 
the Drude peak is essentially gone already for $T=0$.          
The resistivity can therefore be considered
as ``saturated'' already at $T=0$. The upper limit for the resistivity
(Eq. (\ref{eq:de11})) has, however, such a strong $T$ dependence 
that the term ``saturation'' becomes meaningless in this case
(see Fig. \ref{fig:fig7}). The reason for this strong $T$ dependence 
is both the small band width and the LE coupling to the intramolecular
phonons. The $T$ dependence of the TM model is weaker, due to the larger 
band width and to the HI coupling. Even in this case, however, $\rho(T)$ 
does not become a constant because of the $T$ dependence of $\alpha$ 
and $\gamma$.  

\section{Anderson metal-insulator transition and Mott's
minimum conductivity.} \label{sec:i}   

In the semiclassical treatment of the phonons, the phonons cause a 
static disorder, and the problem is therefore related to 
conduction in disordered system. Thus the LE and HI couplings 
correspond to diagonal and off-diagonal disorder, respectively. 
While the disordered systems are usually studied for small $T$, 
we are here interested in the large $T$ behavior. In the semiclassical 
treatment of the phonons, however, apart from causing disorder, 
$T$ only enters via the Fermi-functions (Eq. (\ref{eq:reb6})), which 
is not important for the qualitative behavior. 

Diagonal disorder can lead to an Anderson metal-insulator
transition at $T=0$ (Lee and Ramakrishnan, 1985). For 
off-diagonal disorder, however, Antoniou and Economou (1977)             
found that there is no metal-insulator transition if
the Fermi energy is located in some finite region around
the middle of the band. The semiclassical calculations
agree with these results, i.e., localization is found for
LE but not for HI coupling as $T$ is increased.

In the QMC calculation there is no sign of localization 
for LE coupling, just a lack of saturation.  This is natural. 
Localization depends sensitively on the phase factors, which 
are not destroyed by the elastic scattering in a disordered 
system. The phase information is, however, lost in the inelastic 
scattering by phonons at finite $T$, and localization is not 
expected (Lee and Ramakrishnan, 1985). These effects 
is properly included in the QMC but not in the 
semiclassical treatment of the phonons, and therfore localization 
shows up in the semiclassical (Calandra and Gunnarsson, 2002) but 
not in the QMC treatment (Gunnarsson and Han, 2000).  

Mott (1974) has argued that as the disorder increases, there is 
a discontinuous transition from a metal to an insulator at $T=0$.  
He therefore introduced the concept of the minimum conductivity
\begin{equation}\label{eq:ree1}
\sigma_{\rm min}=0.026 {e^2\over \hbar d},
\end{equation}
where $d$ is the nearest neighbor atomic distance. Later work showed
that the transition from a metal to an insulator actually 
is continuous, but that $\sigma_{\rm min}$ is still relevant  
for low but nonzero temperatures (Lee and Ramakrishnan, 1985).
We therefore make a comparison of $\sigma_{\rm min}$  to the 
resistivity in the TM and C$_{60}$ models. Converting Eq. 
(\ref{eq:ree1}) to a resistivity, we obtain
\begin{equation}\label{eq:ree2}
\rho_{\rm max}=1.6 d \ \ {\rm m}\Omega{\rm cm},
\end{equation}
where $d$ is measured in \AA.
Based on experiment, Mott deduced a somewhat larger minimum conductivity
for systems containing transition metal atoms, 
resulting in the maximum resistivity
\begin{equation}\label{eq:ree3}
\rho_{\rm max}=1 \ \ {\rm m}\Omega{\rm cm}.
\end{equation}
   
Mott derived his result for diagonal disorder. His result can most
naturally be compared with our saturation resistivity for HI coupling
(off-diagonal disorder), since saturation is most pronounced in this 
case. The resistivity $\rho_{\rm max}$ is much larger than the saturation 
resistivity obtained above (Eq. (\ref{eq:reb8}, \ref{eq:rec4})) for the 
TM model with a five-fold degenerate orbital ($N_d=5$). For a fcc lattice 
and a half-filled semi-elliptical band it takes the form
\begin{equation}\label{eq:ree4}
\rho_{\rm sat}={0.14 d \over N_d} \ \ {\rm m}\Omega{\rm cm},
\end{equation}
which is of the order of 0.1-0.2 m$\Omega$cm. The corresponding 
conductivity is substantially larger than Mott's minimum conductivity.

\section{Conclusions}\label{sec:j}
We have reviewed experiments showing resistivity saturation, i.e.,
$\rho(T)$ growing more slowly than $\rho(T)\sim T$ for large $T$. 
Resistivity saturation is found for several classes of metals with 
large resistivities, in particular for many transition metal 
compounds. Saturation often happens in such a way that the Ioffe-Regel 
condition, $l\gtrsim d$, remains fulfilled. Over 
the last 15 years, however, a number of metals have been found 
for which the Ioffe-Regel condition is violated. Some of these 
metals probably show resistivity saturation, but at a much larger
value than the Ioffe-Regel resistivity, while in other cases
little or no sign of saturation is seen.  

We have reviewed early theories, presented at a time
when saturation appeared to be universal. Several of the theories
derived saturation. These theories emphasized different
mechanisms for saturation, and no consensus was reached about
which mechanism dominates.             

Here we argue that it is useful to study the problem
using the f-sum rule. By assuming that the (Drude) peak at 
$\omega=0$ is gone and that only incoherent contributions are left, 
we obtain an approximate upper limit, $\rho_{\rm sat}$, to the resistivity, 
which usually has a weak $T$ dependence. Saturation then happens if 
the resistivity initially grows so rapidly that $\rho_{\rm sat}$ is
reached for small values of $T$ and if $\rho_{\rm sat}$ has a 
weak $T$ dependence.    

We have considered three models with qualitatively different 
behavior: 1) A model of weakly correlated transition metal 
compounds, which shows saturation in agreement with the 
Ioffe-Regel condition, 2) a model of strongly correlated 
high-$T_c$ cuprates, which can give saturation but at much
larger values than the Ioffe-Regel resistivity, and 3) a 
model of alkali-doped C$_{60}$ compounds, which shows no 
saturation. 

To derive the Ioffe-Regel condition for model 1)
we assumed i) noninteraction electrons and ii) $T\ll W$.  
Assumption i) is violated for model 2) and assumption ii)
for model 3). The type of electron-phonon coupling, i.e., 
coupling to the level energies or the hopping integrals,
is also important when comparing models 1) and 3). 

We have focused on work where either the electron-phonon or the 
electron-electron interaction was treated alone.  In many cases 
this may be an oversimplification. For instance, the electron-electron 
interaction is believed to be important for alkali-doped fullerides 
(Gunnarsson, 1997), although only the electron-phonon interaction was 
considered here. Generally, the electron-electron interaction should 
reduce the magnitude of the hopping energy and tend to increase the 
resistivity, as was found for the high-$T_c$ cuprates. The interplay 
between the electron-electron and electron-phonon interactions may, 
however, be more intricate, as found for metal-insulator transitions 
(Han {\it et al.}, 2000) and superconductivity (Han {\it et al.}, 2003) 
in alkali-doped fullerides.

\appendix
\section{Mean-free path}
Since the mean-free path and the Ioffe-Regel resistivity play
an important role in the discussion, we give details     
of how these quantities were obtained. The conductivity tensor 
can be written as (Ashcroft and Mermin, 1976)
\begin{equation}\label{eq:2}
{\bf \sigma}=e^2\sum_{\nu}\int {d^3k\over 4\pi^3}\tau_{\nu}({\bf k})
{\bf v}_{\nu}({\bf k}){\bf v}_{\nu}({\bf k})\lbrack -{\partial f\over 
\partial \varepsilon}\rbrack_{\varepsilon=\varepsilon_{\nu}({\bf k})},
\end{equation}
where $f$ is the fermi function, $\tau_{\nu}({\bf k}$) is the relaxation 
time and ${\bf v}_{\nu}({\bf k})$ is the velocity for a state with band 
index $\nu$ and wave vector ${\bf k}$. We assume a 
three-dimensional isotropic system and a spherical Fermi surface 
with one sheet. We furthermore assume that $\tau$ is independent of
$\nu$ and ${\bf k}$. Together with $v_F=\hbar k_F/m$, where $v_F$ is the 
Fermi velocity, and $l=\tau v_F$, this leads to Eq. (1).  Assuming that 
there are $M$ sheets, we find a moderate reduction of the apparent 
mean free path from Eq. (\ref{eq:1}) by a factor $M^{1/3}$. 
For a quasi-two-dimensional system, like the high-$T_c$ cuprates, 
we instead assume a cylindrical Fermi surface with the height 
$2\pi/c$ and radius $k_F$, where $c$ is the average separation 
of the CuO$_2$ planes. Based on photoemission results (Ino {\it et al.}, 
1999; Yoshida {\it et al.}, 2001), we assume a ``large'' Fermi surface 
containing roughly one electron or hole (more precisely $1\pm x$ carriers, 
where $x$ is the doping). This leads to $k_F=\sqrt{2\pi}/a$, 
where $a$ is the lattice parameter of the CuO$_2$ plane. Then
\begin{equation}\label{eq:3}
\rho_{2d}={2\pi \hbar c\over e^2k_F l}.
\end{equation} 
Assuming the Ioffe-Regel condition $l=a$, we find           
\begin{equation}\label{eq:4}
\rho_{2d}^{Ioffe}=0.055 (c/a_0) \hskip0.5cm {\rm m}\Omega{\rm cm},
\end{equation}
where $a_0=0.529$ \AA \ is the Bohr radius and we have used the
conversion $\hbar a_0/e^2=0.022$ m$\Omega$cm. Assuming $c=6.4$ \AA, 
appropriate for La$_{2-x}$Sr$_x$CuO$_4$, we obtain 
$\rho_{2d}^{Ioffe}=0.7$ m$\Omega$cm. If we instead had assumed 
a ``small'' Fermi surface, containing $x$ carriers, the resistivity
would have been $0.7/\sqrt{x}$. 
Even in this case, the experimental
resistivity of La$_{2-x}$Sr$_x$CuO$_4$ in Fig. \ref{fig:2} exceeds 
the Ioffe-Regel resistivity for small $x$.   

For A$_3$C$_{60}$ (A= K, Rb),
the Ioffe-Regel resistivity was calculated by assuming that $l$ 
is the separation of two C$_{60}$ molecules. Scattering inside the 
molecule is not possible at small and intermediate $T$, since this 
would involve scattering into states that are at least 10000 K higher 
in energy. Therefore the intermolecular separation is the appropriate 
length scale.

\parindent -10pt
\vskip0.4cm
{\bf REFERENCES}
\vskip0.4cm

Abraham, J.M., and B. Deviot, 1972, J. Less-Common Metals {\bf 29}, 311.

Allen, P.B., 1980a, in {\it Superconductivity in d- and
f-Band Metals} H. Suhl and M.B. Maple, Eds. (Academic, New York)
p. 291.

Allen, P.B., 1980b, in {\it Physics of Transition Metals}, P. Rhodes, 
Ed. (Inst. Phys. Conf. Ser. No. 55) p. 425.

Allen, P.B., and B. Chakraborty, 1981, Phys. Rev. B {\bf 23}, 4815.

Allen, P.B., R.M. Wentzcovitch, W.W. Schulz, and P.C. Canfield, 1993,
Phys. Rev. B {\bf 48}, 4359.

Allen, P.B., H. Berger, O. Chauvet, L. Forro, T. Jarlborg, A. Junod,
B. Revaz, and G. Santi, 1996, Phys. Rev. B {\bf 53}, 4393.

Antoniou, P.D., and E.N. Economou, 1977, Phys. Rev. B {\bf 16}, 3768.

Arko, A.J., F.Y. Fradin, and M.B. Brodsky, 1973, Phys. Rev. B {\bf 8}, 4104.

Ashcroft, N.W., and N.D. Mermin, 1976 {\it Solid State Physics}
(Holt, Rinehart and Winston, New York), p. 259. 

Bass, J., 1982, in {\it Landolt-B\"ornstein: Numerical data and functional 
relationships in science and technology}, New Series III/15a, edited 
by K.-H. Hellwege and J.L. Olsen, (Springer, Berlin), p. 1. 

Belitz, D., and W. Schirmacher, 1983, J. Phys. C: Solid State Phys. 
{\bf 16}, 913.

Blankenbecler, R., D.J. Scalapino, and R.L. Sugar, 1981, Phys. Rev. D 
{\bf 24}, 2278.

Calandra, M., and O. Gunnarsson, 2001, Phys. Rev. Lett. {\bf 87}, 266601.

Calandra, M., and O. Gunnarsson, 2002, Phys. Rev. B {\bf 66}, 205105.

Calandra, M., and O. Gunnarsson, 2003, Europhys. Lett. {\bf 61},88.

Carrington, A., D. Colson, Y. Dumont, C. Ayache, A. Bertinotti, 
and J.F. Marucco, 1994, Physica C {\bf 234}, 1. 

Chakraborty, B., and P.B. Allen, 1979, Phys. Rev. Lett. {\bf 42}, 736.

Chen, X.H., M. Yu, K.Q. Ruan, S.Y. Li, Z. Gui, G.C. Zhang, and
L.Z. Cao, 1998, Phys. Rev. B {\bf 58}, 14219.

Christoph, V., and W. Schiller, 1984, J.Phys. F: Met. Phys. {\bf 14}, 1173.

Cote, P.J., and L.V. Meisel, 1978, Phys. Rev. Lett. {\bf 40}, 1586.

Daignere, A., A. Wahl, V. Hardy, and A. Maignan, 2001, Physica C 
{\bf 349}, 189.

Degiorgi, L., B. Briceno, M.S. Fuhrer, A. Zettl,
and P. Wachter, 1994, Nature {\bf 369}, 541.

Duan, H.M., R.M. Yandrofski, T.S. Kaplan, B. Dlugosch, J.H. Wang, 
and A.M. Hermann, 1991, Physica C {\bf 185-189}, 1283.

Fisk, Z., and G.W. Webb, 1976, Phys. Rev. Lett.  {\bf 36}, 1084. 

Fisk, Z., and A.C. Lawson, 1973, Solid State Commun. {\bf 13}, 277.

Forro, L., 2002 (priv. comm.).

Girvin, S.M., and M. Jonson, 1980, Phys. Rev. B {\bf 22}, 3583.

Greenwood, D.A., 1958, Proc. Phys. Soc. {\bf 71}, 585.

Grimvall, G., 1981, {\it The electron-phonon interaction 
in metals}, North-Holland (Amsterdam) pp. 210-223.

Grimvall, G., 2001, private commun.

Grimvall, G., M. Thiessen, and A.F. Guillermet, 1987, Phys. Rev. B
{\bf 36}, 7816.

Guillermet, A.F. and G. Grimvall, 1991, Phys. Rev. B {\bf 44}, 4332.

Gunnarsson, O., 1997, Rev. Mod. Phys. {\bf 69}, 575.

Gunnarsson, O., and J.E. Han, 2000, Nature  {\bf 405}, 1027.

Gurvitch, M., A.K. Ghosh, B.L. Gyorffy, H. Lutz, O.F. Kammerer,
J.S. Rosner, and M. Strongin, 1978, Phys. Rev. Lett. {\bf 41}, 1616.
 
Gurvitch, M., and A.T. Fiory, 1987, Phys. Rev. Lett. {\bf 59}, 1337.

Han, J.E., E. Koch, and O. Gunnarsson, 2000, Phys. Rev. Lett. 
{\bf 84}, 1276.

Han, J.E., O. Gunnarsson, and V.H. Crespi, 2003, Phys. Rev. Lett. 
{\bf 90}, 167006.

Hidaka, Y., and M. Suzuki, 1989, Nature {\bf 338}, 635.

Hou, J.G.,  V.H. Crespi, X.-D. Xiang, W.A. Vareka, G. Briceno, 
A. Zettl, and M.L. Cohen, 1993, Solid State Commun. {\bf 86}, 643.

Hou, J.G., L. Lu, V.H. Crespi, X.-D. Xiang, A. Zettl, and M.L. Cohen, 
1995, Solid State Commun. {\bf 93}, 973. 

Imry, Y., 1980, Phys. Rev. Lett. {\bf 44}, 469.

Ino, A., C. Kim, T. Mizokawa, Z.-X. Shen,
A. Fujimori, M. Takaba, K. Tamasaku, H. Eisaki, and S. Uchida, 1999,
J. Phys. Soc. Jpn {\bf 68}, 1496. 

Ioffe, A.F., and A.R. Regel, 1960, Prog. Semicond. {\bf 4}, 237.

Jarrell, M., and J.E. Gubernatis,  1996, Phys. Rep. {\bf 269}, 133.

Jarrell, M., and Th. Pruschke, 1994, Phys. Rev. B {\bf 49}, 1458.

Jonson, M., and S.M. Girvin, 1979, Phys. Rev. Lett. {\bf 43}, 1447.

Hebard, A.F., T.T.M. Palstra, R.C. Haddon, and R.M. Fleming, 1993,
Phys. Rev. B {\bf 48}, 9945.

Khalifah, P., K.D. Nelson, R. Jin, Z.Q. Mao, Y. Liu, Q. Huang,
X.P.A. Gao, A.P. Ramirez, and R.J. Cava, 2001, Nature {\bf 411}, 669.

Kitazawa, K., T. Matsuura, S. Tanaka, 1981, in {\it Ternary 
Superconductors}, edited by G.K. Shenoy, B.D. Dunlap, and F.Y. Fradin,
Elsevier, (North-Holland, New York), p. 83.

Klein, L., L. Antognazza, T.H. Geballe, M.R. Beasley, and 
A. Kapitulnik, 1999a, Phys. Rev. B {\bf 60}, 1448. 

Klein, L., L. Antognazza, T.H. Geballe, M.R. Beasley, and 
A. Kapitulnik, 1999b, Physica B {\bf 259-261}, 431.
 
Kohn, W., and J.M. Luttinger, 1957, Phys. Rev. {\bf 108}, 590.

Korotin, M.A., V.I. Anisimov, D.I Khomskii, and G.A. Sawatzky, 1998,
Phys. Rev. Lett. {\bf 80}, 4305.

Kubo, Y., Y. Shimakawa, T. Manako, T. Kondo, and H. Igarashi, 1991,
Physica C {\bf 185-189}, 1253.

Lange, E., and G. Kotliar, 1999, Phys. Rev. B {\bf 59}, 1800.

Laughlin, R.B., 1982, Phys. Rev. B {\bf 26}, 3479.

Lee, P.A., and T.V. Ramakrishnan, 1985, Rev. Mod.  Phys. {\bf 57}, 287.

Lewis, S.P., P.B. Allen, and T. Sasaki, 1997, Phys. Rev. B {\bf 55}, 10253.

L\"ohle, J., J. Karpinski, A. Morawski, and P. Wachter, 1996, Physica C
{\bf 266}, 104.

Mahan, G.D., 1990, {\it Many-Particle Physics}, Plenum (New York), p. 651.

Mandrus, D., L. Forro, C. Kendziora, and L. Mihaly, 1992,
Phys. Rev. B {\bf 45}, 12640.

Marchenko, V.A., 1973, Sov. Phys.-Solid Sate {\bf 15}, 1261.

Maldague, P.F., 1977, Phys. Rev. B {\bf 16}, 2437.
 
Martin, R., K.R. Mountfield, and L. Corruccini, 1978, J. de Phys. 
(Paris) {\bf 39}, C6-371 (1978).
 
Martin, S., A.T. Fiory, R.M. Fleming, L.F. Schneemeyer, and J.V. 
Waszczak, 1990, Phys. Rev. B {\bf 41}, 846.

Mazin, I.I., D.J. Singh, and C. Ambrosch-Draxl, 1999, J. Appl. Phys.
{\bf 85}, 6220.

Merino, J., and R.H. McKenzie, 2000, Phys. Rev. B {\bf 61}, 7996.

Millis, A.J., J. Hu, and S. Das Sarma, 1999, Phys. Rev.
Lett. {\bf 82}, 2354.

Mooij, J.H., 1973 phys. stat sol. (a) {\bf 17}, 521.

Mott, N.F., 1974, {\it Metal-insulator transitions}, Taylor $\&$ Francis 
(London).

Morton, N., B.W. James, and G.H. Wostenholm, 1978, Cryogenics {\bf 18}, 131.

Orenstein, J., G.A. Thomas, A.J. Millis, S.L. Cooper,
D.H. Rapkine, T. Timusk, L.F. Schneemeyer, and J.V. Waszczak, 1990,
Phys. Rev. B {\bf 42}, 6342.

Parcollet, O., and A. Georges, 1999, Phys. Rev. B {\bf 59}, 5341.

Palstra, T.T.M., A.F. Hebard, R.C. Haddon, and P.B. Littlewood, 1994,
Phys. Rev. B {\bf 50}, 3462.

Pickett, W.E., K.M. Ho, and M.L. Cohen, 1979, Phys. Rev. B {\bf 19}, 1734.

Rodbell, D.S., J.M. Lommel, and R.C. DeVries, 1966, J. Phys. Soc. Japan
{\bf 21}, 2430.

Ron, A., B. Shapiro, and M. Weger, 1986, Phil. Mag. B {\bf 54}, 553.

Ruan, K.Q., Q. Cao, S.Y. Li, G.G. Qian, C.Y. Wang, X.H. Chen, L.Z. Cao,
2001, Physica C {\bf 351}, 402.
 
Salamon, M.B. and M. Jaime, 2001, Rev. Mod. Phys, {\bf 73}, 583. 

Singley, E.J., C.P. Weber, D.N. Basov, A. Barry, and J.M.D. Coey, 1999,
Phys. Rev. {\bf 60}, 4126.

Stephens, P.W., L. Mihaly, P.L. Lee, R.L. Whetten, S.-M. Huang, 
R. Kaner, F. Deiderichs, and K. Holczer, 1991, Nature {\bf 351}, 632.

Sunandana, C.S., 1979, J. Phys. C: Solid State Phys. {\bf 12}, L165.
 
Sundqvist, B., and B.M. Andersson, 1990, Solid State Commun. {\bf 76}, 1019.

Takagi, H., B. Batlogg, H.L. Kao, J. Kwo, R.J. Cava, J.J. Krajewski, 
and W.F. Peck, Jr., 1992, Phys. Rev. Lett. {\bf 69}, 2975.

Takenaka, K, R. Shiozaki, S. Okuyama, J. Nohara, A. Osuka, 
Y. Takayanagi, and S. Sugai, 2002, Phys. Rev. B {\bf 65}, 092405.

Tsuei, C.C., 1986, Phys. Rev. Lett. {\bf 57}, 1943.
 
Tyler, A.W., A.P. Mackenzie, S. NishiZaki, and Y. Maeno, 1998,
Phys. Rev. B {\bf 58}, R10107.

Uchida, S., T. Ido, H. Takagi, T. Arima, Y. Tokura, and S. Tajima, 1991,
Phys. Rev. B {\bf 43}, 7942.

Vareka, W.A., and A. Zettl, 1994, Phys. Rev. Lett. {\bf 72}, 4121.

Wang, N.L., C. Geibel, and F. Steglich, 1996a, Physica C {\bf 262},  231.

Wang, N.L., B. Buschinger, C. Geibel, and F. Steglich, 1996b, 
Phys. Rev. B {\bf 54}, 7445.

Wang, N.L., P. Zheng, T. Feng, C.D. Gu, C.C. Homes, J.M. Tranquada, 
B.D. Gaulin, and T. Timusk, 2002, cond-mat/0211224.

Weger, M., and N.F. Mott, 1985, J. Phys. C: Solid State Phys.
{\bf 18}, L201.

Weger, M., 1985, Phil Mag. B {\bf 52}, 701.

Wiesmann, H., M. Gurvitch, H. Lutz, A. Ghosh, B. Schwartz, M. Strongin, 
P.B. Allen, and J.W. Halley, 1977, Phys. Rev. Lett. {\bf 38}, 782.

Woodard, D.W., and G.D. Cody, 1964, Phys. Rev. {\bf 1964}, A166.

Zhang, F.C., and T.M. Rice, 1988, Phys. Rev. B {\bf 37}, 3759.

Yamada, K.,  C.H. Lee, K. Kurahashi, J. Wada, S. Wakimoto, S. Ueki,
H. Kimura, Y. Endoh, S. Hosoya, G. Shirane, R.J. Birgeneau,
M. Greven, M.A. Kastner, and Y.J. Kim, 1998, Phys.  Rev. B {\bf 57}, 6165.

Yan, S.L., Y.Y. Xie, J.Z. Wu, T. Aytug, A.A. Gapud, B.W. Kang,
L. Fang, M. He, S.C. Tidrow, K.W. Kirchner, J.R. Liu, and W.K. Chu,
1998, Appl. Phys. Lett. {\bf 73}, 2989.

Yoshida, T., X.J. Zhou, M. Nakamura, S.A. Kellar, P.V. Bogdanov, 
E.D. Lu, A. Lanzara, Z. Hussain, A. Ino, T. Mizokawa, A. Fujimori, 
H. Eisaki, C. Kim, Z.-X. Shen, T. Kakeshita, and S. Uchida, 2001, 
Phys. Rev. B {\bf 63}, 220501(R).

\end{multicols}
\end{document}